\documentclass[superscriptaddress, reprint, prl,twocolumn,footinbib,aps,nobalancelastpage]{revtex4-2}

\usepackage[ usenames, dvipsnames ]{xcolor}
\usepackage{amsmath, amsfonts, amssymb, amsthm, amstext, bm, bbm, multirow}
\usepackage[colorlinks=true, linkcolor=blue, citecolor=red, 
urlcolor=blue, anchorcolor = blue, filecolor = blue, hypertexnames=true]{hyperref}
\usepackage{graphicx, subcaption}

\usepackage{ragged2e}
\DeclareCaptionJustification{justified}{\justifying}
\newcommand{\brac}[1]{\left( #1 \right) }
\newcommand{\rmd}[0]{\mathrm{d}} 
\newcommand{\mcj}[0]{\mathbb{J}}

\newcommand{\figref}[1]{Fig.~\ref{fig:#1}}

\begin{document}

\title{Nonequilibrium spin transport in integrable and non-integrable classical spin chains}

\author{Dipankar Roy}
\email{dipankar.roy@icts.res.in}
\affiliation{International Centre for Theoretical Sciences, Tata Institute of Fundamental Research, Bangalore 560089, India}
\author{Abhishek Dhar}
\email{abhishek.dhar@icts.res.in}
\affiliation{International Centre for Theoretical Sciences, Tata Institute of Fundamental Research, Bangalore 560089, India}
\author{Herbert Spohn}
\email{spohn@ma.tum.de}
\affiliation{Zentrum Mathematik and Physik Department, Technische Universität München, Garching 85748, Germany}
\author{Manas Kulkarni}
\email{manas.kulkarni@icts.res.in}
\affiliation{International Centre for Theoretical Sciences, Tata Institute of Fundamental Research, Bangalore 560089, India}

\date{\today}

\begin{abstract}
Anomalous transport in low dimensional spin chains is an intriguing topic that can offer key insights into the interplay of integrability and symmetry in many-body dynamics. Recent studies have shown that spin-spin correlations in spin chains, where integrability is either perfectly preserved or broken by symmetry-preserving interactions, fall in the Kardar-Parisi-Zhang (KPZ) universality class. Similarly, energy transport can show ballistic or diffusive-like behaviour. Although such behaviour has been studied under equilibrium conditions, no results on nonequilibrium spin transport in classical spin chains has been reported so far. In this work, we investigate both spin and energy transport in classical spin chains (integrable and non-integrable) when coupled to two reservoirs at two different temperatures/magnetization. In both the integrable case and broken-integrability (but spin-symmetry preserving), we report anomalous scaling of spin current with system size ($\mcj^s \propto L^{-\mu}$) with an exponent, $\mu \approx 2/3$, falling under the KPZ universality class. On the other hand, it is noteworthy that energy current remains ballistic ($\mcj^e \propto L^{-\eta}$ with $\eta \approx 0$) in the purely integrable case and there is departure from ballistic behaviour ($\eta > 0$) when integrability is broken regardless of spin-symmetry. Under nonequilibrium conditions, we have thoroughly investigated spatial profiles of local magnetization and energy. We find interesting nonlinear spatial profiles which are hallmarks of anomalous transport. We also unravel subtle striking differences between the equilibrium and nonequilibrium steady state through the lens of spatial spin-spin correlations.

\end{abstract}

\maketitle

% -------------------------
% introduction 
% -------------------------
Anomalous behaviour in low dimensional systems is a fascinating phenomenon with great significance in theory and applications. Recent discoveries on such phenomena that deviates from Fourier law in one-dimensional, integrable spin chains has generated a lot of interest \cite{2021-bulchandani--ilievski, 2023-gopalakrishnan-vasseur-feb, 2023-gopalakrishnan-vasseur-may}. It has been observed that the well-known \emph{quantum Heisenberg} spin-$\frac{1}{2}$ chain (also called quantum \textit{XXX} model) exhibits anomalous behaviour for nonequilibrium spin transport \cite{2011-znidaric, 2017-ljubotina--prosen}. In particular, this anomalous behaviour involves deep connections to the Kardar-Parisi-Zhang (KPZ) universality class \cite{1986-kardar--zhang, 2018-takeuchi}, wherein spin correlation is characterized by the KPZ scaling function \cite{2004-prahofer--spohn, 2019-ljubotina--prosen}. Such anomalous behaviour, often referred to as the \emph{KPZ superdiffusion}, also exists in other integrable quantum models with higher symmetry \cite{2020-dupont-moore, 2020-krajnik--prosen}. Moreover, there is numerical evidence that the KPZ superdiffusion is present in all integrable models having non-Abelian symmetry \cite{2022-ye--yao}. There has been significant progress in the experimental side as well, with contemporary experiments detecting the 1D KPZ physics in real systems modelled by the quantum Heisenberg spin-$\frac{1}{2}$  model \cite{2021-scheie--tennant, 2022-wei--zeiher}. To understand such behaviour in the quantum models, a number of analytical studies have focused on the relation between integrability and the KPZ superdiffusion \cite{2018-ilievski--prosen, 2019-gopalakrishnan-vasseur, 2021-ilievski--ware, 2021-bulchandani--ilievski}. In addition to these results for integrable models, more recent studies have considered the impact of additional integrability-breaking terms. Weak integrability-breaking, spin-symmetry preserving terms leave the KPZ superdiffusion intact in the quantum Heisenberg chain \cite{2021-nardis--ware}. However, normal diffusion with enhanced diffusion coefficient is observed for spin in the quantum Heisenberg model at an infinite temperature in the presence of strong noisy couplings \cite{2022-claeys--arbeitman}. The question of integrability breaking in quantum systems is still open because various parameter regimes are yet to be understood \cite{2021-bastianello--vasseur, 2022-nandy--prelovsek, 2023-gopalakrishnan-vasseur-feb, 2023-gopalakrishnan-vasseur-may}.

Remarkably, even in classical spin chains KPZ superdiffusion has been reported. Numerical investigations in Refs.~\onlinecite{2013-prosen-zunkovic, 2019-das--dhar, 2020-krajnik-prosen} show that the KPZ superdiffusion describes spin correlations in the \emph{integrable lattice Landau-Lifshitz} (ILLL) also known as the \emph{Ishimori-Haldane} (IH) model \cite{1982-ishimori, 1982-haldane}. Studying classical analogues are specially beneficial because of severe numerical challenges in quantum models.

 In a recent study motivated by the quantum-classical correspondence \cite{2019-das--dhar, 2020-nardis--vasseur, 2020-richter--steinigeweg, 2021-schubert--steinigeweg, 2022-heitman--steinigeweg}, robustness of the KPZ superdiffusion is predicted in the classical integrable spin chain in the presence of integrability-breaking but spin-symmetry preserving terms \cite{2023-roy--kulkarni}. When spin-symmetry is not preserved, either diffusive or anomalous (but non-KPZ) behaviour arises. Interestingly, a special situation occurs at low temperatures even in the \emph{classical Heisenberg} (CH) model which is non-integrable: KPZ superdiffusion is observed due to near-integrability \cite{2022-mcroberts--moessner}. At intermediate temperatures, the possibility of anomalous behaviour is debatable \cite{2013-bagchi, 2019-li, 2022-mcroberts--moessner}. 
 
 %In the presence of strong disorder, the CH model shows energy diffusion \cite{2009-oganesyan--huse}. 

Most of these aforementioned studies rely on the equilibrium correlations to glean information about transport properties. The number of studies on spin or energy transport under \emph{nonequilibrium} conditions are limited. On the quantum side, there has been some work on diffusive energy and/or spin transport \cite{2005-monasterio--casati, 2009-prosen-znidaric, 2011-znidaric}. On the classical side, thermal tranpsort has been investigated in the CH model \cite{2007-savin--zotos, 2012-bagchi-mohanty}. It is worth noting that there has been no work reported on nonequilibrium spin transport in classical spin chains even for the well-known CH model.

Before proceeding further, we recall some important aspects of nonequilibrium energy (or heat) transport which has been a topic of intense research \cite{2000-bonetto--bellet, 2003-lepri--politi, 2008-dhar, 2016-lepri, 2020-benenti--livi, 2020-iubini--politi, 2022-livi, 2023-benenti--livi}.  Early studies on heat conduction and temperature profile in microscopic models include the theoretical works on harmonic chains, both regular and disordered, in the nonequilibrium steady state (NESS) \cite{1967-rieder--lieb, 1970-matsuda-ishii, 1971-casher-lebowitz, 1974-connor-lebowitz}. These studies observe diverging conductivity (with system size) in harmonic chains. Such divergence in the thermodynamic limit is also observed in the Fermi-Pasta-Ulam-Tsinghou (FPUT) model (see \cite{2007-gallavotti} for a review) where anharmonic interactions are present \cite{1997-lepri--politi, 1998-lepri--politi, 2007-mai--narayan}. There are also one-dimensional systems of particles with alternating masses which exhibit anomalous transport \cite{1999-hatano, 2001-dhar, 2002-grassberger--yang, 2003-casati-prosen, 2005-cipriani--politi, 2007-mai--narayan}. Moreover, there has been investigations in other low-dimensional systems, such as, the discrete nonlinear Schr\"{o}dinger equation \cite{2013-iubini--politi, 2015-borlenghi--fransson-apr, 2015-borlenghi--fransson-jul}, the 1D Toda chains \cite{1979-toda, 1999-hatano, 2010-iacobucci--stoltz}, rotor models \cite{2016-iubini--politi}, nanosystems \cite{2023-benenti--livi}, to name a few. To obtain an understanding of anomalous heat transport starting from a microscopic model, there has been analytical studies on simple models as well, for example, the stochastic model in Refs.~\cite{2009-lepri--politi, 2010-delfini--politi}, the L\'{e}vy walk model \cite{2013-dhar--derrida}, and the harmonic chain momentum exchange model \cite{2019-kundu--dhar}. 

%\textcolor{red}{Stop.}

\begin{figure}[htbp!]
	\centering
	\includegraphics[width=1\linewidth]{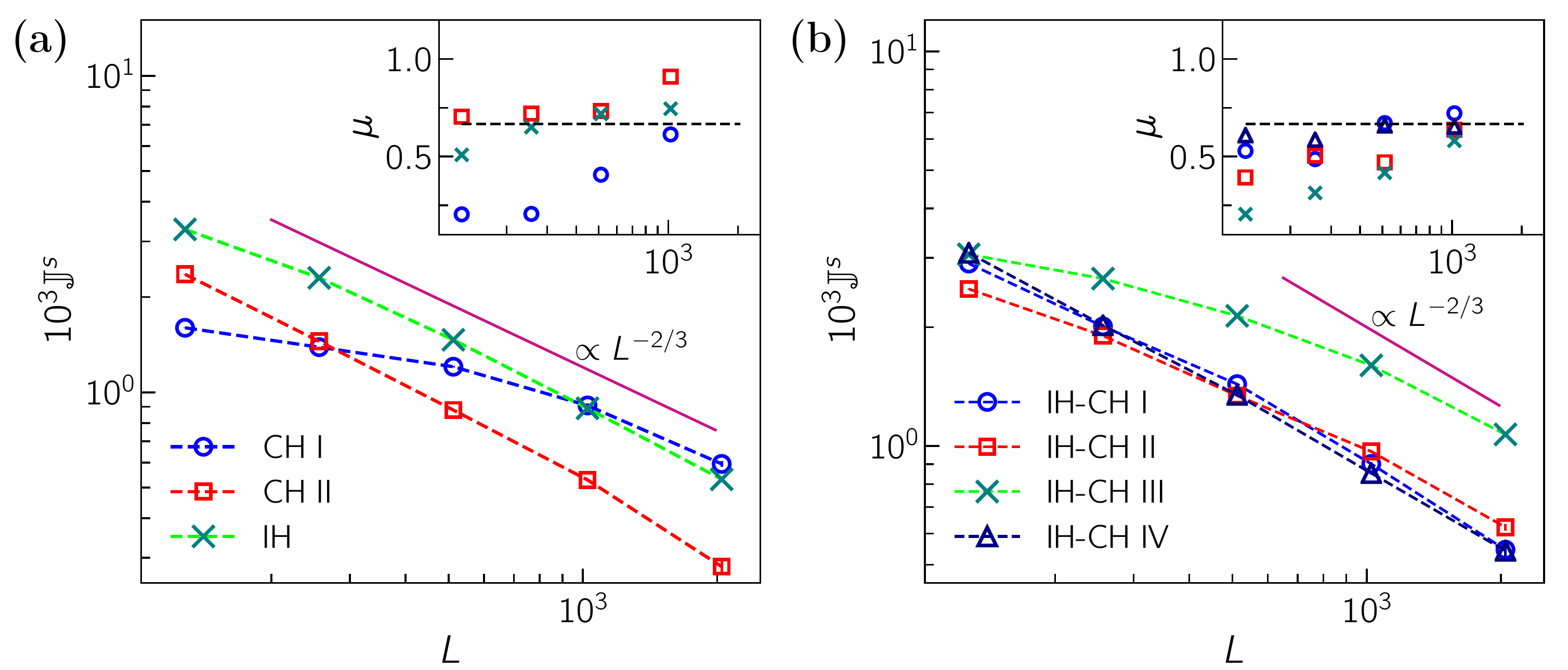}
	\caption{Plots of spin current $\mcj^s$ [Eq.~\eqref{eq:curr}] versus system size $L$ for (a) the IH and CH models, and (b) the IH-CH spin chain. The error bars (computed using the average bond currents) are smaller than the symbols for all data-points in both the plots and hence not visible. We average over $48-50$ independent simulations over $4$ different time intervals of length $ \mathcal{T}_{av} = 10^5 $ in the NESS. The two insets show the exponent $\mu$ computed from two neighbouring data points. The horizontal dashed line in the insets correspond to $\mu = 2/3$ (KPZ superdiffusion).}
	\label{fig:jsvl}
\end{figure}

\begin{figure}[htbp!]
		\centering
		\includegraphics[width=1\linewidth]{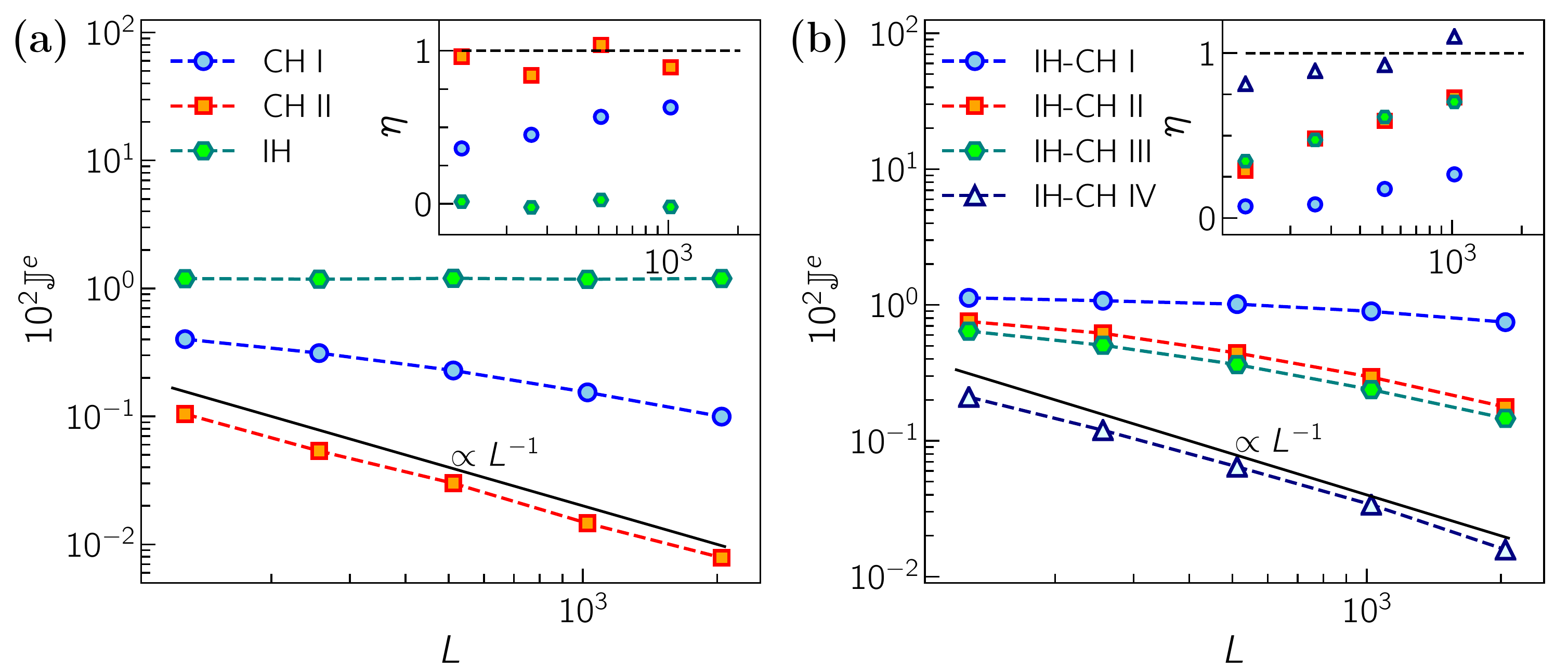}
		\caption{Plots of the energy current $\mcj^e$ [Eq.~\eqref{eq:curr}] versus system size $L$ for (a) the CH and IH models, and (b) the IH-CH spin chain. The error bars (computed using the average bond currents) are smaller than the symbols for all data-points in both the plots. We average over $48-50$ independent simulations over $4-5$ different time intervals of length $ \mathcal{T}_{av} = 10^5 $ in the NESS. The two insets show the exponent $\eta$ computed from two neighbouring data points. The horizontal dashed line in the insets correspond to $\eta = 1$ (conventional diffusion).}
		\label{fig:jevl}
	\end{figure}

Given this enormous interest about anomalous transport in low-dimensional systems, we investigate the spin and energy current in classical integrable and non-integrable spin chains. It is to be noted that even in the non-integrable model there are four conservation laws, namely, the energy and three components of spin (due to isotropy). The key findings of our work are as follows: 
\begin{enumerate}

    \item[(i)] \textit{Integrable case (IH):} We observe that spin current scales with system size as $L^{-\mu}$ with $\mu \approx 2/3$ in the integrable IH model [see \figref{jsvl}(a)]. This is a hallmark of KPZ superdiffusion. The energy current however is observed to be ballistic [see \figref{jevl}(a)] thereby bringing out the signatures of integrability.

    \item[(ii)] \textit{Non-integrable case (CH):} In the non-integrable CH model, for the spin current, we find hints of KPZ superdiffusion at low temperature and conventional diffusion at high temperature [see \figref{jsvl}(a)]. On the other hand, for energy, we find the energy current approaches conventional diffusive behaviour [see \figref{jevl}(a)] thereby consistent with the expectation from generic non-integrable/chaotic systems.

    \item[(iii)] \textit{Broken-integrability case (IH-CH):} This case pertains to a case where the IH chain is perturbed with a non-integrable (CH) Hamiltonian. Remarkably, we find that the spin current scaling with system size falls into the KPZ universality class even in the non-perturbative regime [see \figref{jsvl}(b)].
    
\end{enumerate}

%We devise a novel numerical scheme suitable for studying both spin and energy transport under nonequilibrium conditions.

% -------------------------
% models and definitions
% -------------------------

\begin{figure}[htbp!]
	\includegraphics[width=1\linewidth]{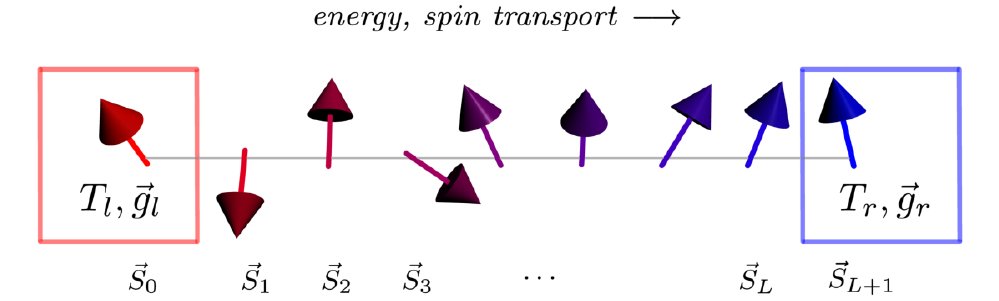} %sc-plot.pdf}
	\caption{(Color online) A schematic diagram showing the 1D classical spin chain ($\vec{S}_1 , \vec{S}_2, \ldots, \vec{S}_L$) in contact with two reservoirs represented by $\vec{S}_0$ and $\vec{S}_{L+1}$. We introduce two auxiliary fields $\vec{g}_l$ and $\vec{g}_{r}$ to ensure that the boundary spins $\vec{S}_0$ and $\vec{S}_{L+1}$ are maintained at a desired magnetization.}
	\label{fig:schem}
\end{figure}

The model we consider (\figref{schem}) is a classical spin chain of $L+2$ three-component spins, $\vec{S}_n, n=0,1,2,\ldots, L+1 $, of unit length.  The Hamiltonian of the system is 
\begin{equation}
	\begin{aligned}
		\mathcal{H} & = \sum_{n=0}^{L} e_{n} ,  \\
		e_n 		& = - J  \ln \! \left( 1 + \vec{S}_n \cdot \vec{S}_{n+1} \right) - \lambda  \vec{S}_n \cdot \vec{S}_{n+1} 
	\end{aligned} 
	\label{eq:ham}
\end{equation}
with $J, \lambda \geqslant 0$ describing the relative strength of the integrable and non-integrable parts. We refer to this model as the \emph{Ishimori-Haldane-classical-Heisenberg} or IH-CH spin chain. Note that $e_n$ in Eq.~\eqref{eq:ham} is local energy. It is also to be noted that $\lambda=0$ corresponds to the IH spin chain or the ILLL model \cite{1982-haldane, 1982-ishimori, 1988-theodorakopoulos, 1987-papanicolaou, 2019-das--dhar} and $J=0$ corresponds to the CH spin chain \cite{1967-joyce}.

{In order to study nonequilibrium spin and energy transport, in addition to the Hamiltonian dynamics, we incorporate local heat bath dynamics at the boundary spins 
$\vec{S}_0$ and $\vec{S}_{L+1}$  
~(see Supplementary Material~\cite{2023-supp} for details). The local heat bath dynamics of the two boundary spins is specified by their temperatures $T_l$ and $T_r$ and by two auxiliary boundary magnetic fields $\vec{g}_l$ and $\vec{g}_{r}$. 
 The two boundary spins thus  play the role of reservoirs, which means that the  bond energy of the pair of spins $\vec{S}_0$ and $\vec{S}_1$, and the magnetization of $\vec{S}_0$  are set by the  temperature $T_{l}$ and field $\vec{g}_l$.  Likewise,
 the  bond energy of the pair of spins $\vec{S}_L$ and $\vec{S}_{L+1}$, and the magnetization of $\vec{S}_{L+1}$  are set by the  temperature $T_{r}$ and field $\vec{g}_r$. }
 Our full dynamics consists of alternating periods of Hamiltonian evolution of the $L+2$ spins and Monte-Carlo evolution of the two boundary spins~\cite{2023-supp}.

%$\vec{g}_0$ and $ \vec{g}_{N+1}$ acts on the spins $\vec{S}_0$ and $\vec{S}_{N+1}$ respectively. 

%The dynamics of the bulk spins in the IH-CH model is given by
%\begin{equation}
%    \frac{\mathrm{d} \vec{S}_n }{ \mathrm{d} t } = \vec{S}_n \times \brac{ c_{n-1} \vec{S}_{n-1} +     c_n \vec{S}_{n+1} },
%\end{equation} 
%where 
%$
%	c_n = \lambda  + J / \brac{ 1 + \vec{S}_n \cdot \vec{S}_{n + 1} }.
%$
%Thus, we can rewrite this equation as 

The dynamics of the bulk spins in the IH-CH model is given by
\begin{equation}
    \frac{\rmd \vec{S}_n }{ \rmd t } 
    = 
    - \brac{ \vec{\mcj}^{s}_n - \vec{\mcj}^{s}_{n-1} } ,
\end{equation} 
where we define local \emph{spin} current as $\vec{\mcj}^{s}_n = - c_n \,   ( \vec{S}_n \times \vec{S}_{n+1} )$ with 
$
	c_n = \lambda  + J / ( 1 + \vec{S}_n \cdot \vec{S}_{n + 1} ).
$ Similarly, we can write an equation for local energy $e_n$ 
\begin{equation}
    \frac{\rmd e_n }{ \rmd t } 
    = 
    - \brac{ \mcj^{e}_n - \mcj^{e}_{n-1} } ,
\end{equation}
where local \emph{energy} current $\mcj^{e}_n = - c_n \vec{S}_n \cdot \vec{\mcj}^{s}_{n+1} $. We consider the lattice-averaged currents $\mcj^{s}$ and $\mcj^{e}$ defined as 
\begin{equation}
    \mcj^{s} = \frac{1}{L-2} \sum_{n=1}^{L-2} {\Big\langle} \vec{\mcj}^{s}_n \cdot \hat{z} {\Big\rangle}, \qquad
    \mcj^{e} = \frac{1}{L-2} \sum_{n=1}^{L-2} {\Big\langle} \mcj^{e}_n {\Big\rangle},
    \label{eq:curr}
\end{equation}
where the average $\langle \cdot \rangle$ is over time (in NESS) as well as different realizations and $\hat{z}$ is the unit vector corresponding to the third component. Note that the lower and upper limit of the summation in Eq.~\eqref{eq:curr} is chosen so as to ensure that $\vec{S}_0$ and $\vec{S}_{N+1}$ do not contribute to the currents. We expect that spin and energy currents scale with system size as 
\begin{equation}
    \mcj^{s} \propto L^{-\mu},  \quad \mcj^{e} \propto L^{-\eta} ,  \qquad \mu, \eta \geqslant 0.
\end{equation}
We focus on determining the exponents $\mu$ and $\eta$ in our numerical simulations (see supplementary material \cite{2023-supp}). It is worth pointing out that for some models, the theoretically best understood cases are ballistic, KPZ, and diffusive  for which the exponent equals $0,2/3$, and $1$
respectively.

%\begin{widetext}
\begin{center}
\begin{table}[htbp!]
	\renewcommand*{\arraystretch}{1.6}
	\begin{center}
		\begin{tabular}{|c|c|c|c|c|c|c|}
			\hline 
			Model  & Case & \ \ $J$ \ \ & \ \ $\lambda$ \ \ & \ \ $T$ \ \ & \ \quad $\mu $ \quad \ & \ \quad $\eta $ \quad \ \\ \hline
			\multirow{2}{*}{ \ \ CH \ \ } & I  & $0$ & $1$ & 0.2 & $0.61 $ & $0.63$ \\ \cline{2-7}
			& II  &  $0$ & $1$ & 1 & $0.91$ & $0.89$ \\ \hline
			\multirow{1}{*}{IH/ILLL } &  & $  1  $ & $0$ & $ 1 $ & $0.75$ & $-0.02$  \\ \hline
			\multirow{4}{*}{IH-CH}& I & $1$ & $0.1$ & $1$ & $0.72$ & $0.26$ \\ \cline{2-7}
			& II & $1$ & $0.5$ & $1$ & $0.64$ & $0.72$ \\ \cline{2-7}
			& III & $1$ & $1.5$ & $1$ & $0.58$ & $0.70$ \\ \cline{2-7}
			& IV & $0.2$ & $1$ & $1$ & $0.65$ & $1.10$ \\ \cline{2-7}
			\hline
		\end{tabular}
	\end{center}
	\vspace{0.2cm}	
	\caption{ The models and corresponding values of the parameters $J$ and $ \lambda$ considered in this study. The corresponding exponents $\mu$ and $\eta$ for spin and energy transport respectively from our simulations are listed. $T$ here is average temperature of the two reservoirs, i.e. $T = (T_l + T_r ) / 2$. 
 The column ``Case'' labels the parameter set.   
        } 
	\label{table:models}
\end{table}
\end{center}

\smallskip

% In order to check that our protocol suits the study of both spin and energy transport, we consider two cases for the boundary spins: (1) no field, i.e. $g=0$, and (2) a nonzero field with $g=0.5$ for the CH and IH models. For these checks, we fix $T=1$ and do not apply any temperature gradient, i.e. $\Delta T = 0$. We compute the following correlation at the centre of the system:
% \begin{equation}
% 		C_{ob}(x) = \langle \vec{S}_{N/2} \cdot \vec{S}_{N/2 + x} \rangle_{ss} .
% \end{equation}  
% We compute $C(x)$ under the open boundary condition (OBC) with the above parameters using the protocol mentioned above. For comparison, we also consider periodic boundary condition (PBC) with a uniform field of strength $g$ along the $z$-axis at each site. As in the case of OBC, we consider two values of $g$: $g=0$ and $g=0.5$. We use Monte Carlo algorithm to generate equilibrium configurations and then compute lattice-averaged correlation $C_{pb}(x) = \sum_{i=1}^{N}\langle \vec{S}_{i} \cdot \vec{S}_{i+x} \rangle_{eq } / N$. We plot these correlations in Fig.~\ref{fig:corr} and observe that the correlations $C_{pb}(x)$ and $C_{ob}(x)$ are in good agreement with each other.

%\begin{widetext}
%\onecolumngrid

%\twocolumngrid
%\end{widetext}

%\begin{widetext}
%\onecolumngrid

%\twocolumngrid
%\end{widetext}

% ------------------------- 
% results
% ------------------------- 

We consider seven cases (see Table~\ref{table:models}), i.e. points in the $\brac{J,\lambda,T}$ space, for our IH-CH model to investigate spin and energy transport. One can distinguish two broad categories for these seven cases on the basis of the parameters $J$ and $\lambda$: (a) when $J=1, \lambda =0$ or $J=0, \lambda = 1$ which correspond to purely integrable (IH) or purely non-integrable (CH) respectively, and (b) $J, \lambda > 0 $ which represent the mixed case where integrable Hamiltonians are coupled to terms resulting in breaking of integrability, but preserving spin-symmetry. The figures presented in this paper are divided in these two categories as well. Other parameters [$T_l, T_r, \vec{g}_l=(0,0,g), \vec{g}_{r}=(0,0,-g)$] that are used in our simulations are described in detail in the Supplementary Material \cite{2023-supp}. Below we discuss our findings.

%The values of the parameters $T_b$, $\Delta T$, and $g$ used in our simulations are provided in Table~\textcolor{blue}{I} in the Supplemental Material \cite{2023-supp}. 

%\medskip\noindent
\textit{CH model (CH I and CH II). --} The nature of spin transport depends on the temperature in the CH model. In particular, we observe KPZ superdiffusion at low temperatures where the scaling of spin current is $\mcj^s \sim L^{-\mu}$ with $\mu \approx 0.61 $ at $T=0.2$ [see \figref{jsvl}(a) and inset]. This finding of KPZ nonequilibrium transport is rooted to the possible deep connection between low temperature limit of CH model and continuum Landau-Lifshitz or IH model \cite{2022-mcroberts--moessner}. On the other hand, in high temperature regime, we find close to diffusive behaviour. At $T=1$, we obtain an exponent $\mu \approx 0.91$. For the energy current, at low temperature ($T=0.2$), we find that $\mcj^e \sim L^{-\eta}$ with $\eta\approx 0.63$. However, we see a trend that suggests that the exponent is likely to increase as $L$ is increased [see \figref{jevl}(a) and inset]. At high temperature ($T=1$), we find close to conventional diffusion for the energy current with the scaling exponent $\eta \approx 0.89$.

%, but either anomalous (non-KPZ) or diffusive behaviour is exhibited at higher temperatures \cite{2022-mcroberts--moessner}. We find  and $\mu \approx 0.91$ at $T=1$ [see Fig.~\ref{fig:jsvl}(a)]. These values are close to the expected values which are $\mu=2/3$ at $T=0.2$ and $\mu=1$ at $T=1$. Energy transport is diffusive at high temperatures \cite{2012-bagchi-mohanty, 2022-mcroberts--moessner}. Similar to Ref.~\onlinecite{2012-bagchi-mohanty}, we observe close to $\sim 1/L$ behaviour for energy currents. The exponent based on the last two data points is found to be $\eta \approx 0.89$ [see \figref{jevl}(a)]. At low temperature, the exponent is $\eta \approx 0.63$ [see \figref{jevl}(a)] which suggests anomalous behaviour in contrast to diffusive behaviour observed in \cite{2022-mcroberts--moessner}. It is possible that energy diffusion is observed if we consider larger $L$. We also compute the correlation $C_{neq}(n)$ defined as
%\begin{equation}
%    C_{neq}(n) =  \langle \vec{S}_{N/2} \cdot \vec{S}_{N/2 + n} \rangle_{ss}  - \langle \vec{S}_{N/2} \rangle_{ss}  \cdot \langle \vec{S}_{N/2 + n} \rangle_{ss}.
%\end{equation}
%for (1) $g=0$ and (2) $g=0.5$ (see \figref{neqcor}). We observe that the correlation is exponentially decreasing for $g=0$ in agreement with the results for equilibrium correlations. However, the correlation does not decay for $g=0.5$.  

%\medskip\noindent
\textit{IH model. --} Departure from Fourier law are expected for both spin and energy transport in this model because of its integrability. We find that the spin current scales as $\mcj \sim L^{-\mu}$ with $\mu \approx 0.72$ [see \figref{jsvl}(a) and inset]. This KPZ scaling of spin current with system size is indeed remarkable and consistent with scaling exponents obtained from equilibrium correlations \cite{2019-das--dhar}. On the other hand, the energy transport is observed to be ballistic with the energy current scaling as $\sim L^{-\eta}$ with $\eta \approx -0.02$ [see \figref{jevl}(a) and inset]. It is worth noting that having spin current scaling with system size to be superdiffusive and energy current to be ballistic in the same model is rather rare and remarkable. 

Having discussed the purely integrable ($J=1, \lambda =0$) and the purely non-integrable ($J=0, \lambda =1$) cases so far, next we discuss the mixed cases ($J, \lambda > 0$).

%\medskip\noindent
\textit{Weakly perturbative regime (IH-CH I). --} 
First, we set $\lambda=0.1$ and $J=1$ such that integrability is broken with a weak perturbation. In this weakly perturbative regime, we observe an exponent $\mu \approx 0.72$ [see \figref{jsvl}(b) and inset] which indicates a behaviour close to KPZ superdiffusion. On the other hand, for energy current, we find that the exponent is $\eta \approx 0.26$ as shown in \figref{jevl}(b). This behaviour is due to finite system size and upon increasing the system size we expect that $\eta$ would tend to $1$ (conventional diffusion).

\medskip\noindent
\textit{Intermediate regime (IH-CH II). --}
We increase the strength of the perturbation by setting $\lambda=0.5$ keeping $J=1$. In this intermediate regime, where energy contribution from the integrable and nonintegrable terms are comparable, we also observe the KPZ behaviour for spin transport with $\mu \approx 0.64$ [see \figref{jsvl}(b) and inset]. The closeness to KPZ exponent even when the perturbation is considerable is remarkable. On the other hand, the energy current scales with system size with an exponent $\eta \approx 0.73 $ for energy [see \figref{jevl}(b) and inset]. This again is a finite size effect and upon increasing system size we expect $\eta$ to increase to $1$ (conventional diffusion).

%\medskip\noindent
\textit{Nonperturbative regime (IH-CH III). --}
To access the nonperturbative regime, we increase the perturbation strength to $\lambda=1.5$. Here also, we find evidence of close to KPZ behaviour for spin transport with the exponent $\mu \approx 0.58$ [see \figref{jsvl}(b) and inset]. Just like in the previous case, we obtain an exponent $\eta \approx 0.73 $ for energy current [see \figref{jevl}(b) and inset]. The exponent is expected to increase to $1$ (conventional diffusion) when the system size is increased.

\medskip\noindent
\textit{Other perturbative regime (IH-CH IV). --}
In addition to the three cases mentioned above (IH-CH I, IH-CH II, IH-CH III), we also check a case where the integrable part plays the role of perturbation. We achieve this by setting $J=0.2$ and $\lambda = 1$. We find $\mu \approx 0.65$ for the spin current [see \figref{jsvl}(b) and inset]. This suggests the KPZ superdiffusion for spin transport. We also compute the energy currents for different system sizes and find an exponent $\eta \approx 1.10$ [see \figref{jevl}(b) and inset]. This indicates diffusive energy transport.

\begin{figure}[htbp!]
	\begin{subfigure}{1\linewidth}
		\centering
		\includegraphics[width=1\linewidth]{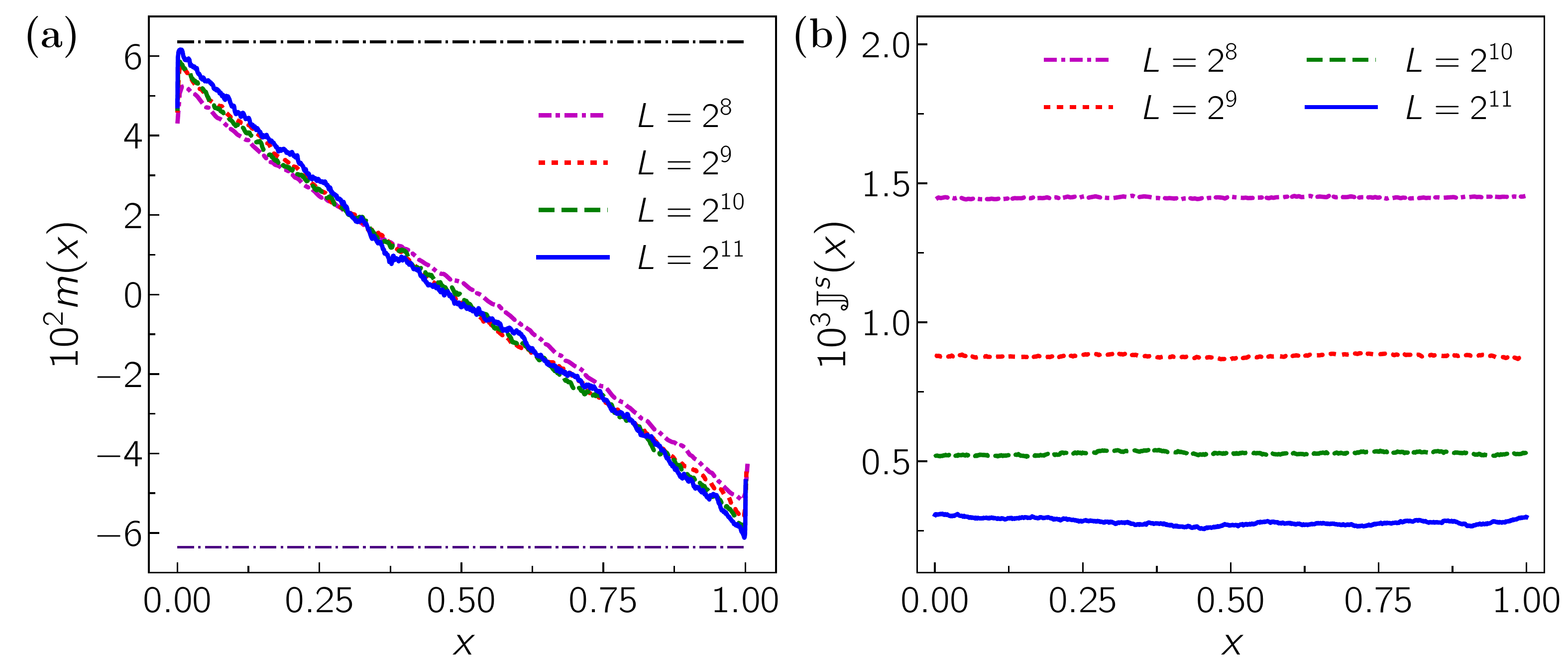}
	\end{subfigure}
	\begin{subfigure}{1\linewidth}
		\centering
		\includegraphics[width=1\linewidth]{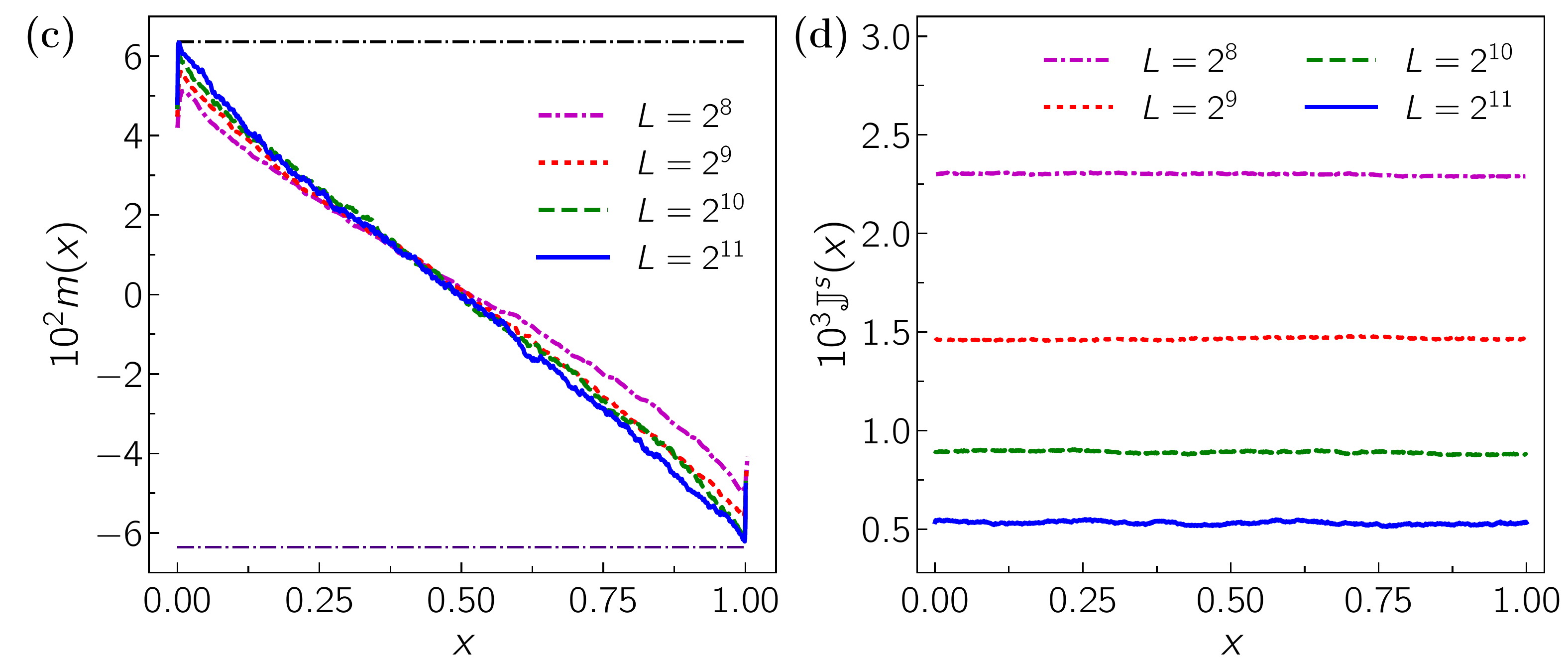}
	\end{subfigure}
	\caption{ (a) Plot of average magnetization $m(x) = \langle S^{z}_{ [xL] } \rangle$ for the CH model (CH II) where we have chosen $T=1$. We see that the magnetization profile is linear which is a fingerprint of conventional diffusive behaviour. The horizontal dashed-dotted lines show expected values of magnetizations at the boundaries. The top horizontal line shows the expected magnetization at the left boundary ($\vec{S}_0$) while the bottom line shows the expected magnetization at the right boundary ($\vec{S}_{L+1}$). (b) Plot of average bond current $\mathbb{J}^{s}(x)$ versus normalized site position $x$ for the CH model. This plot shows that we have reached NESS. (c) Average magnetization $m(x) = \langle S^{z}_{ [xL] } \rangle$ for the IH model where we have chosen $T=1$. Nonlinear magnetization profile is a hallmark of anomalous transport. As in (a), the horizontal dashed-dotted lines show expected values of magnetizations at the boundaries. (d) Average bond current $\mathbb{J}^{s}(x)$ versus normalized site position $x$ for IH model which confirms that we have reached NESS.}
	\label{fig:magprof}
\end{figure}

\begin{figure}[htbp!]
	\centering
	\includegraphics[width=0.9\linewidth]{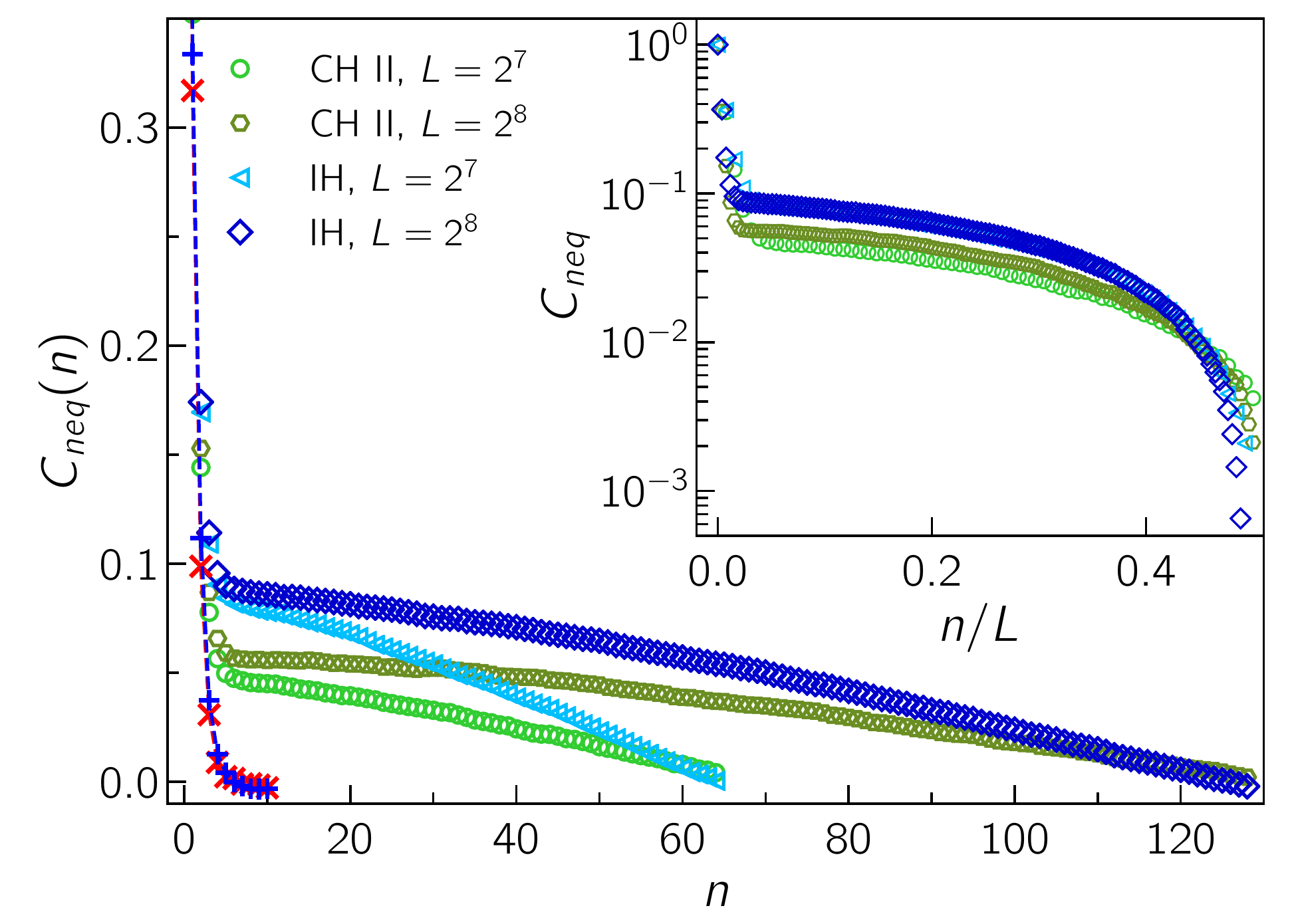}
	\caption{Plots of correlations $C_{neq}(n)$ [Eq.~\eqref{eq:cneq}] for the CH II and IH spin chains at $T=1$ with no field [$g=0$, therefore equilibrium]. Red cross and blue plus denote equilibrium data for CH (II) and IH models respectively. The nonequilibrium case with nonzero field ($g=0.5$) is shown for two system sizes as indicated in the legend. We set $T_l = T_r =1$. We see a stark difference between equilibrium ($\vec{g}_l = \vec{g}_{r}$) and nonequilibrium ($\vec{g}_l \neq \vec{g}_{r}$) situation. We choose system size $L=256$ for the equilibrium case ($g=0$). In the nonequilibrium case, we present two different system sizes ($L=128$ and $L=256$) to show the finite size effects, that are absent in the equilibrium case.  In the inset, we plot $C_{neq}$ versus $n/L$ which shows a  good data-collapse. }
	\label{fig:neqcor}
\end{figure}

Having discussed results on nonequilibrium spin and energy transport in detail, we now discuss an important quantity namely spatial profiles of average magnetization when the spin chain is placed out of equilibrium. In \figref{magprof}, we present results for the average magnetization $m(x) = \langle S^{z}_{ [xL] } \rangle$ where $[xL] $ equals the integer closest to $xL$ and recall that $\langle \cdot \rangle$ denote average over time (in NESS) as well as different realizations. We see nonlinear spatial profiles in the IH case [\figref{magprof}(c)] which is the hallmark of anomalous behaviour \cite{2010-chaudhuri--spohn, 2019-dhar--kundu} and linear profiles in the CH case [\figref{magprof}(a)] which is expected in conventional diffusive systems. It is to be kept in mind that nonlinear profiles can also occur in conventional diffusion equation with a density dependent diffusion constant but the difference in the anomalous case  is that the nonlinearity persists for arbitrarily small temperature (or field) difference applied at the two ends.

In addition to local spatial profiles discussed above, it is interesting to investigate spatial correlations in NESS. To do so, we investigate a quantity $C_{neq}(n)$ given by
\begin{equation}
    C_{neq}(n) =  \langle \vec{S}_{ \frac{L}{2} } \cdot \vec{S}_{\frac{L}{2} + n} \rangle  - \langle \vec{S}_{\frac{L}{2}} \rangle  \cdot \langle \vec{S}_{\frac{L}{2} + n} \rangle.
    \label{eq:cneq}
\end{equation}
In \figref{neqcor}, we plot this correlation for the IH and CH models and demonstrate stark difference between equilibrium and nonequilibrium situation. In particular, in the equilibrium case, we have exponentially decaying correlation (short ranged), but in the nonequilibrium case we see long-range correlations. The long-range correlations and their scaling form (as seen in the inset of \figref{neqcor}) is  one of the hallmarks of the NESS~\cite{1990-garrido--spohn} and has also been observed in systems with anomalous transport~ \cite{2019-kundu--dhar}. We also benchmark our results with known equilibrium results (see Supplemental Material \cite{2023-supp}).

% -------------------------
% conclusion
% -------------------------

In conclusion, we have considered purely integrable (IH) and purely non-integrable (CH) spin chains along with perturbed integrable cases while maintaining spin-symmetry (IH-CH). We primarily focus on system size scaling of spin and energy currents and this is summarized in Table~\ref{table:models}. Notably, we find strong evidence of KPZ superdiffusion in nonequilibrium spin transport in (i) IH case, (ii) CH case at low temperature, and (iii) IH-CH cases. To the best of our knowledge, this is the first work reporting spin transport in classical spin chains. We also demonstrate the existence of nonlinear spatial profiles of magnetization which is fingerprint of anomalous behaviour (\figref{magprof}). We compute spatial correlations and remark on the striking difference between the equilibrium and nonequilibrium cases (\figref{neqcor}).

In future, it is interesting to develop an analytical framework presumably based on effective fractional diffusion equation \cite{2019-dhar--kundu} for these various class of spin chains considered in our work. Exploring the role of disorder \cite{2009-oganesyan--huse} and additional spin-symmetry preserving terms \cite{2020-dupont-moore} is interesting both from theoretical and experimental perspective. The presence of anisotropy parameter that violate spin-symmetry preservation \cite{2011-znidaric, 2023-roy--kulkarni} is expected to lead to superdiffusion-diffusion crossover and this is an interesting direction to explore.

H.S., A.D. and M.K. acknowledge the support from the Science and
Engineering Research Board (SERB, government of India),
under the VAJRA faculty scheme (No. VJR/2019/000079).
M.K. would like to acknowledge support from the project 6004-1 of the Indo-French Centre for the Promotion of Advanced Research (IFCPAR), SERB Early Career Research Award (ECR/2018/002085) and SERB Matrics Grant (MTR/2019/001101) from the Science and Engineering Research Board (SERB), Department of Science and Technology (DST), Government of India. A.D. and M.K., and acknowledge support of the Department of Atomic Energy, Government of India, under Project No. 19P1112RD. M.~K. thanks the hospitality of LPENS (Paris), LPTHE (Paris) and LPTMS (Paris-Saclay).  
%use our protocol to study nonequilibrium transport in the presence of disorder, additional spin-symmetry preserving terms, and anisotropy. The KPZ superdiffusion is expected to be stable even in the presence of disorder in the integrability broken case. Regular diffusion or anomalous behaviour can appear depending on the anisotropy \cite{2023-roy--kulkarni}. Observation of these transport properties in NESS are exciting future problems. Also, it will interesting to study if the anomalous transport in the spin chains admits a stochastic fractional equation description \cite{2019-kundu--dhar}.  

%\bibliographystyle{apsrev4-1}
%\bibliographystyle{plainnat}
\bibliography{sneq-refs}

\end{document}

\newcommand{\titlename}{\underline{\textsc{Supplemental material}}\\ \bigskip Nonequilibrium spin transport in integrable and non-integrable classical spin chains}

\title[]{\titlename}

\author{Dipankar Roy}
\email{dipankar.roy@icts.res.in}
\affiliation{International Centre for Theoretical Sciences, Tata Institute of Fundamental Research, Bangalore 560089, India}
\author{Abhishek Dhar}
\email{abhishek.dhar@icts.res.in}
\affiliation{International Centre for Theoretical Sciences, Tata Institute of Fundamental Research, Bangalore 560089, India}
\author{Herbert Spohn}
\email{spohn@ma.tum.de}
\affiliation{Zentrum Mathematik and Physik Department, Technische Universität München, Garching 85748, Germany}
\author{Manas Kulkarni}
\email{manas.kulkarni@icts.res.in}
\affiliation{International Centre for Theoretical Sciences, Tata Institute of Fundamental Research, Bangalore 560089, India}
\date{\today}

%\onecolumngrid
%\begin{center} 
%	\textbf{ \textit{\textcolor{red}{\huge -- Incomplete --} } }
%\end{center}
%\twocolumngrid

\maketitle

\tableofcontents 

%\vspace{0.8cm}

% -------------------------
% simulation protocol
% -------------------------

\section{Simulation details}
\label{sec:sim-det}

\begin{figure}[htbp!]
	\includegraphics[width=1\linewidth]{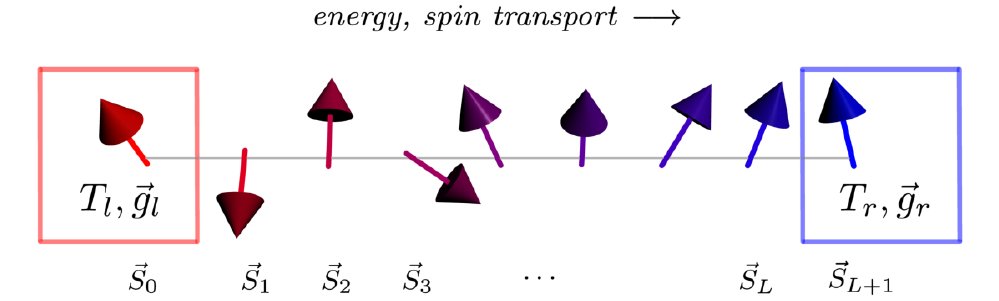} %sc-plot.pdf}
	\caption{(Color online) We recall the schematic diagram showing the 1D classical spin chain ($\vec{S}_1 , \vec{S}_2, \ldots, \vec{S}_L$) in contact with two reservoirs represented by $\vec{S}_0$ and $\vec{S}_{L+1}$. We introduce two auxiliary fields $\vec{g}_l$ and $\vec{g}_{r}$ to ensure that the boundary spins $\vec{S}_0$ and $\vec{S}_{L+1}$ are maintained at a desired magnetization. The detailed simulation procedure is described in Section~\ref{sec:sim-det}.}
	\label{fig:schem_supp}
\end{figure}

In this section, we discuss the direct numerical simulation (DNS) of the nonequilibrium steady state (NESS) for the open spin chain. We recall the schematic  diagram in Fig.~\ref{fig:schem_supp}. 
It is necessary to choose an appropriate protocol for the dynamics of the boundary spins $\vec{S}_0$ and $\vec{S}_{L+1}$ such that these act as effective reservoirs. 
%It is possible that some protocols which work for energy transport may be unsuitable for studying spin transport. To address this, 
Our prescription involves a protocol which is a combination of (i) purely deterministic dynamics of the bulk spins, $\vec{S}_1, \ldots, \vec{S}_{L}$ and (ii) a mixed dynamics (deterministic and Monte Carlo) of the boundary spins, $\vec{S}_0$ and $\vec{S}_{L+1}$. We will discuss the details below. Recall that the Hamiltonian is given by
\begin{equation}
	\begin{aligned}
		\mathcal{H} & = \sum_{n=0}^{L} e_{n} ,  \\
		e_n 		& = - J  \ln \! \left( 1 + \vec{S}_n \cdot \vec{S}_{n+1} \right) - \lambda  \vec{S}_n \cdot \vec{S}_{n+1} 
	\end{aligned} 
	\label{eq:ham_app}
\end{equation}
with $J, \lambda \geqslant 0$ describing the relative strength of the integrable and non-integrable parts.

\begin{figure}[htbp!]
	\centering
	\includegraphics[width=1\linewidth]{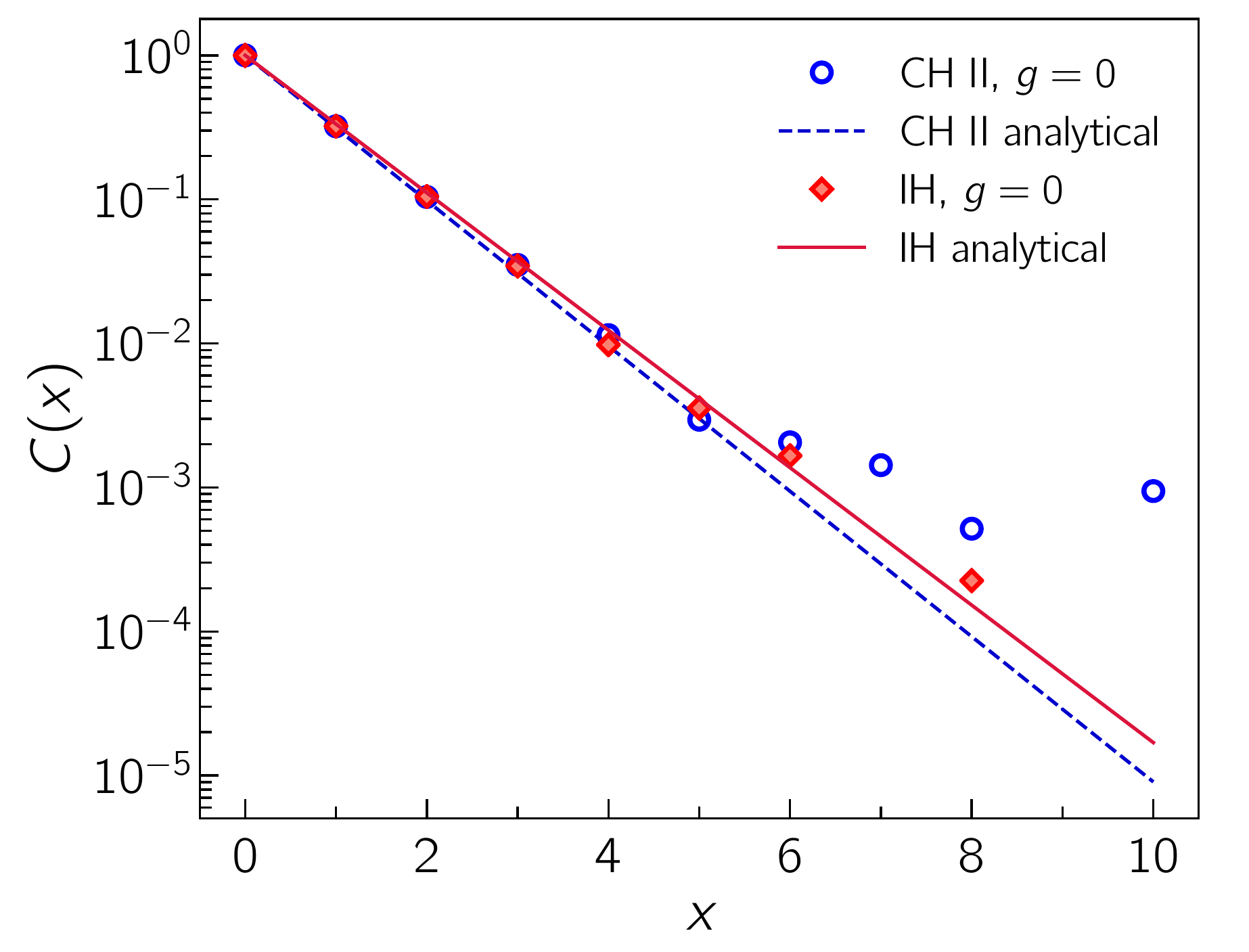}
	\caption{Plots of correlations $C_{pb}(x)$ [Eq.~\eqref{eq:cpb}] for the classical Heisenberg (CH II) and the Ishimori-Haldane (IH) spin chains at $T=1$ with no field ($g=0$). The dashed blue and the solid red line represent analytical results for the CH II and IH models respectively given in Eqs.~\eqref{eq:ch-analytical} and \eqref{eq:ih-analytical} respectively. Here the system size used for numerics is $L=32$.}
	\label{fig:coreq}
\end{figure}

\begin{figure}[htbp!]
    %\begin{center}
    \begin{subfigure}{1\linewidth}
	\includegraphics[width=1.0\linewidth]{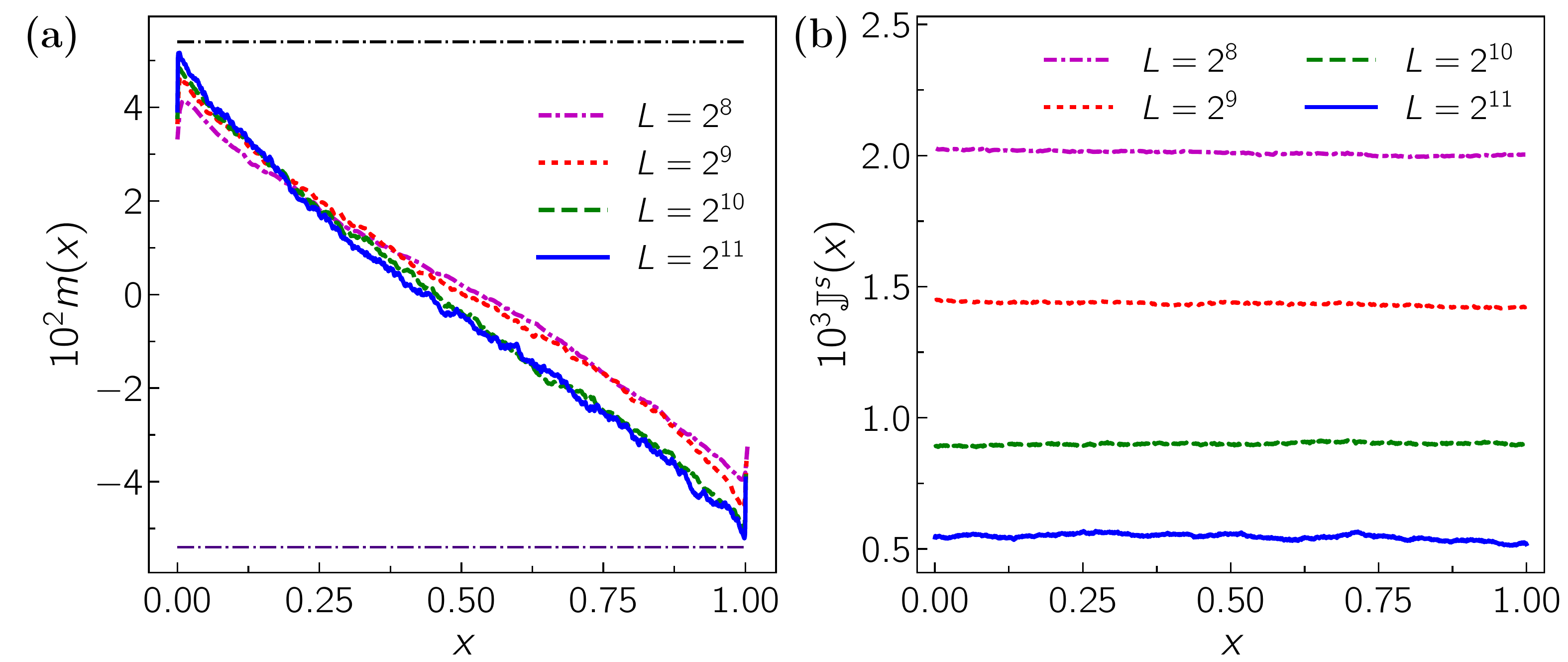}
    \end{subfigure}
    \begin{subfigure}{1\linewidth}
	\includegraphics[width=1.0\linewidth]{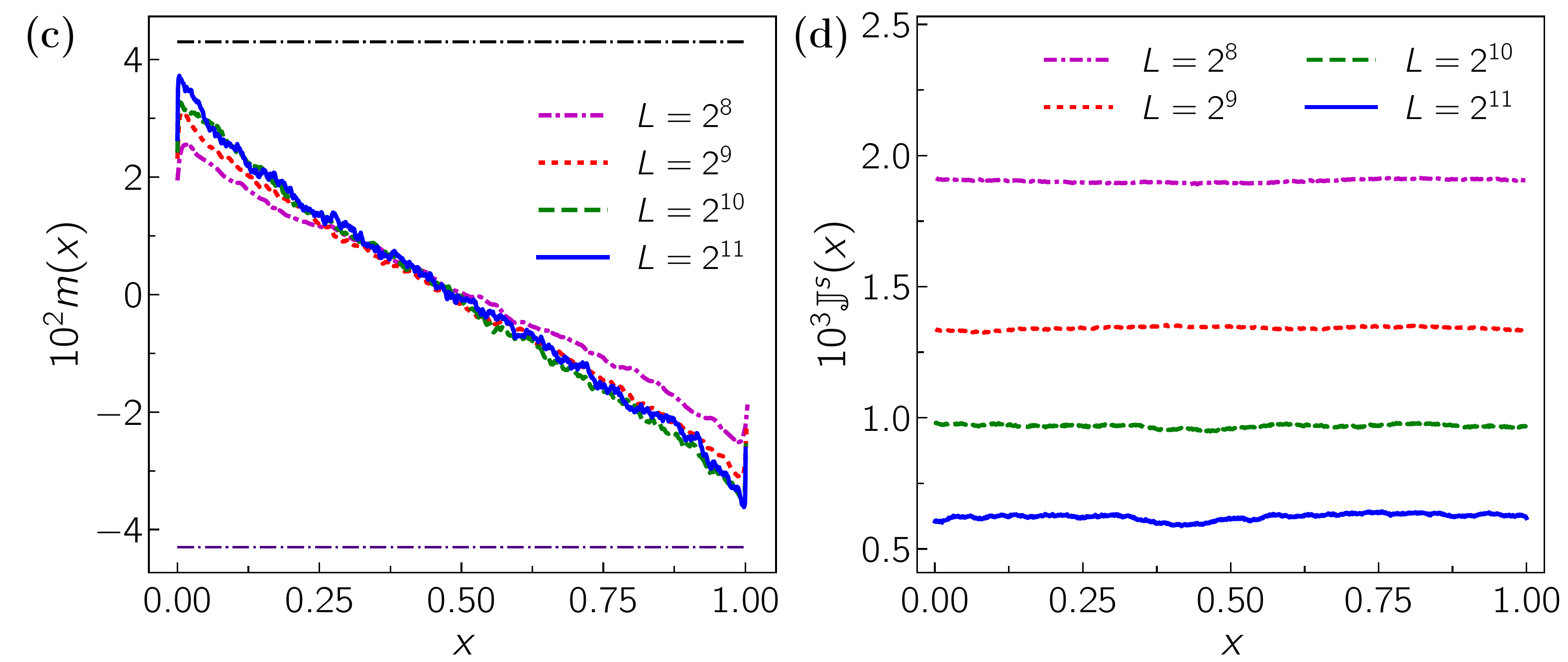}
    \end{subfigure}
    \begin{subfigure}{1\linewidth}
	\includegraphics[width=1.0\linewidth]{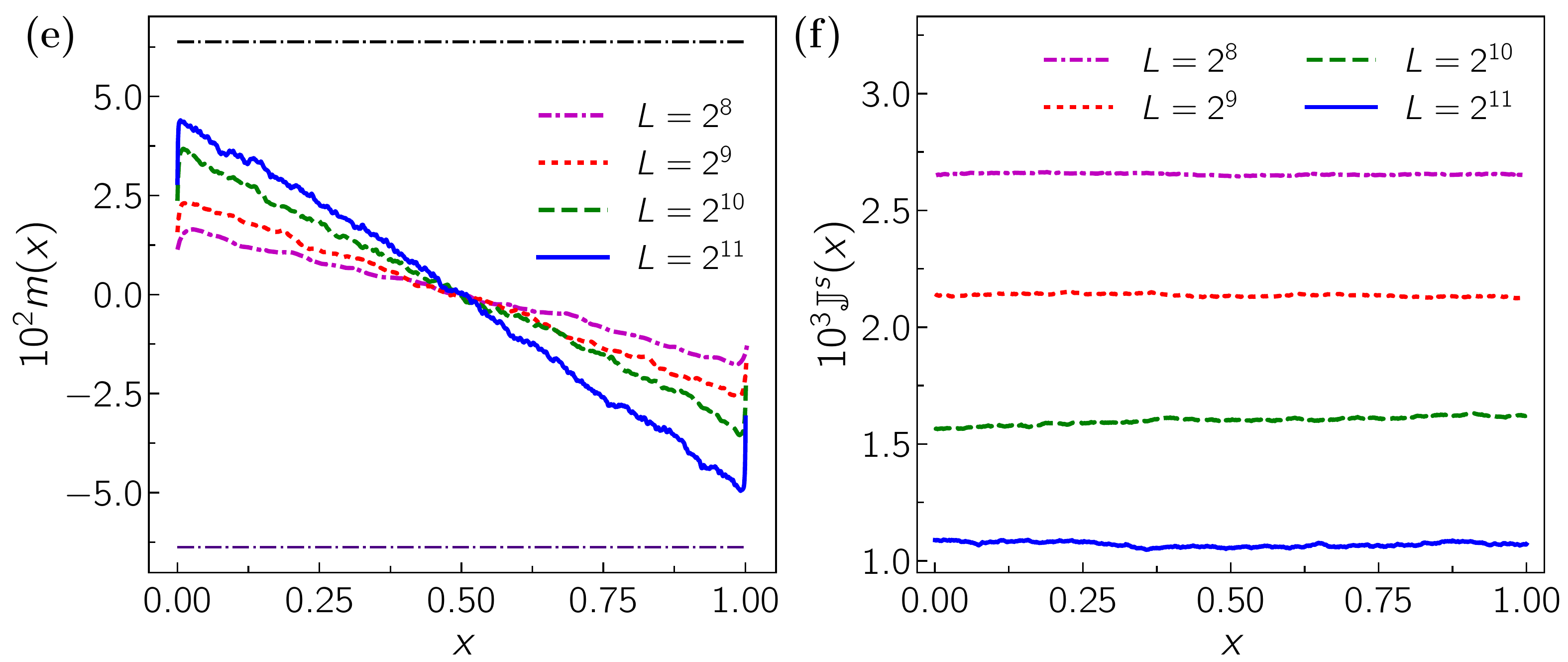}
    \end{subfigure}
    \begin{subfigure}{1\linewidth}
	\includegraphics[width=1.0\linewidth]{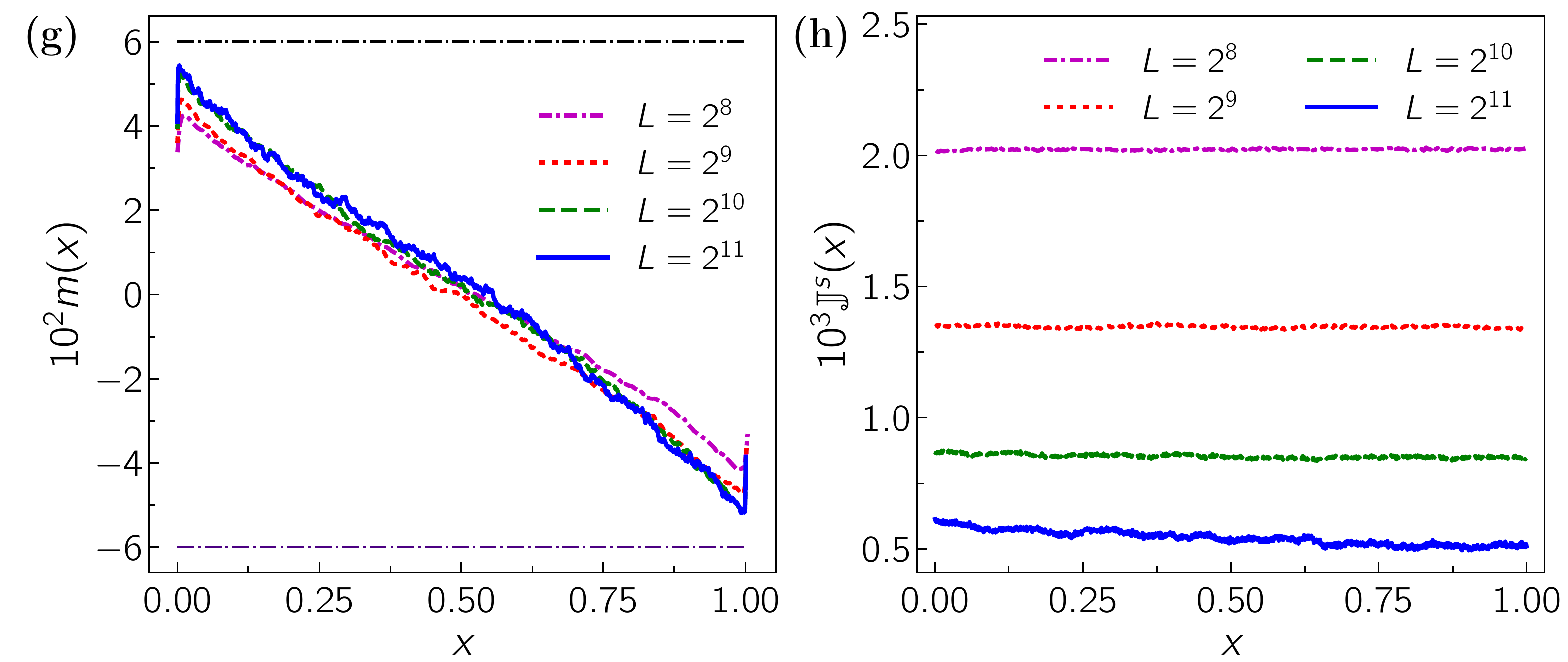}
    \end{subfigure}
    %\end{center}
    \caption{Plots of magnetization profiles in (a), (c), (e), and (g) for the integrability-broken cases IH-CH I, II, III, and IV respectively. The parameters for all these cases are provided in Table~\ref{tab:params}. The temperature is fixed at $T=1$. In all these plots, we observe nonlinear-shaped profiles which is a hallmark of anomalous transport. The horizontal dashed-dotted lines show expected values of magnetizations at the boundaries. In other words, the top horizontal line shows the expected magnetization at the left boundary ($\vec{S}_0$) while the bottom line shows the expected magnetization at the right boundary ($\vec{S}_{L+1}$). In order to benchmark the numerical procedure and to ensure we have truly reached NESS in the right column [(b), (d), (f), and (h)], we have shown the local average bond spin current. The flat profiles confirm that we have reached NESS.
    %as well as spatial profiles for spin current in (b), (d), (f), and (h) for integrability-broken cases IH-CH I, II, III, and IV respectively. The magnetization profiles are nonlinear in all cases for sufficiently large system sizes. The current profiles are flat suggesting that we are in the NESS. }
    }
    \label{fig:spro-ib}
\end{figure}

\begin{figure}[htbp!]
	%\begin{center}
	\begin{subfigure}{1\linewidth}
		\includegraphics[width=1.0\linewidth]{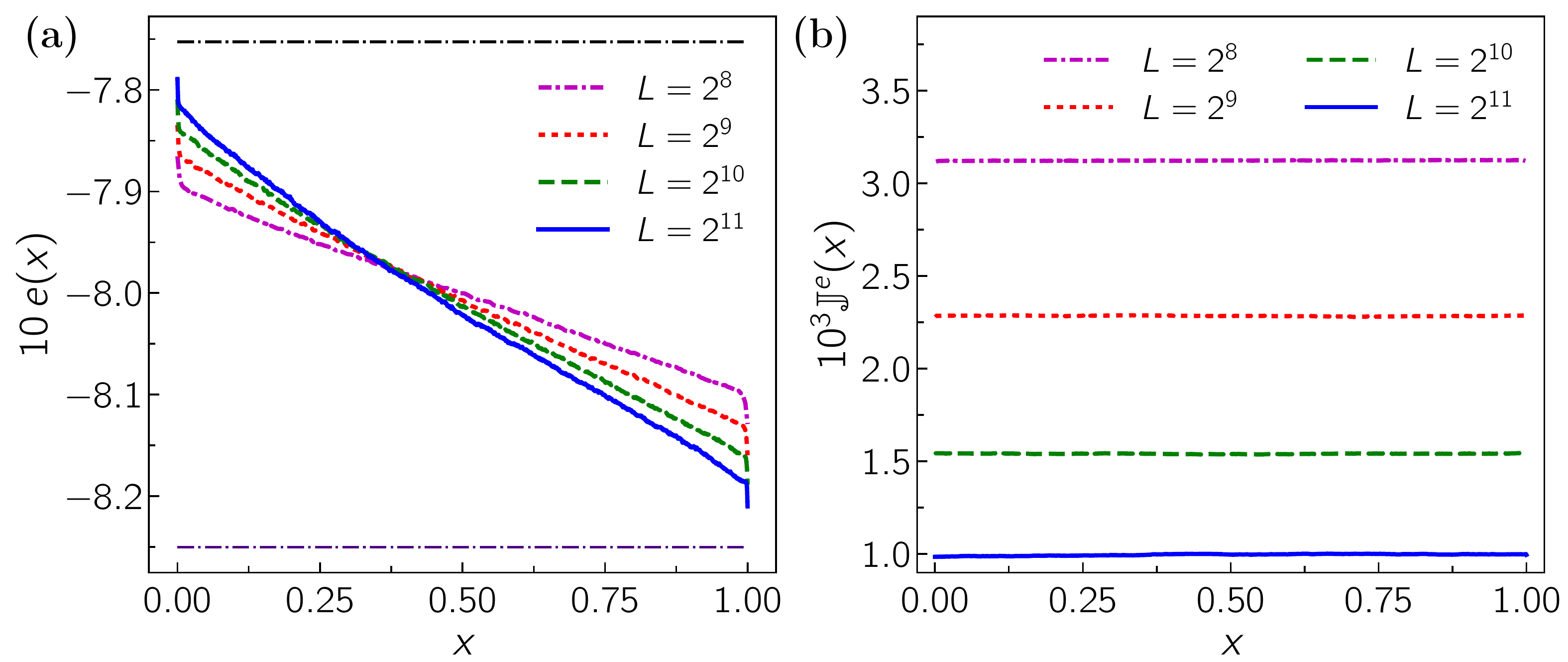}
	\end{subfigure}
	\begin{subfigure}{1\linewidth}
		\includegraphics[width=1.0\linewidth]{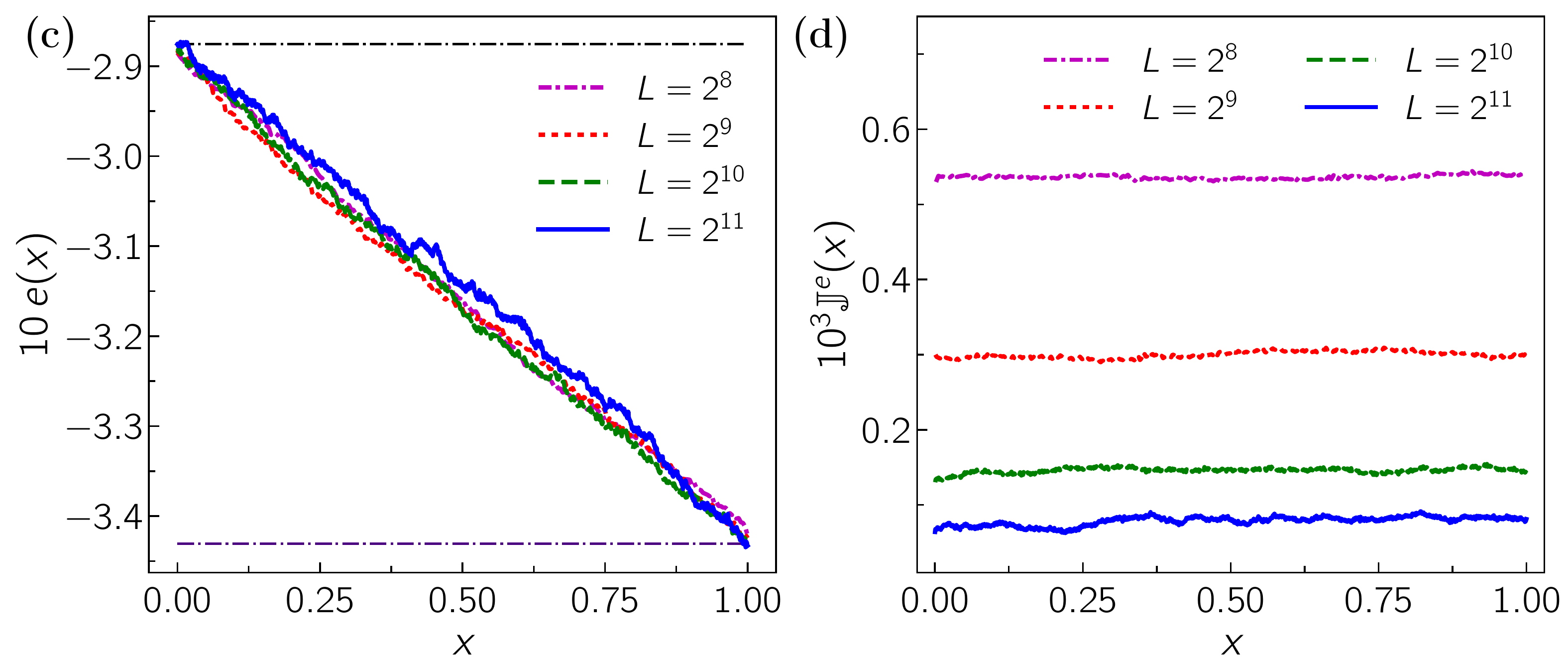}
	\end{subfigure}
	\begin{subfigure}{1\linewidth}
		\includegraphics[width=1.0\linewidth]{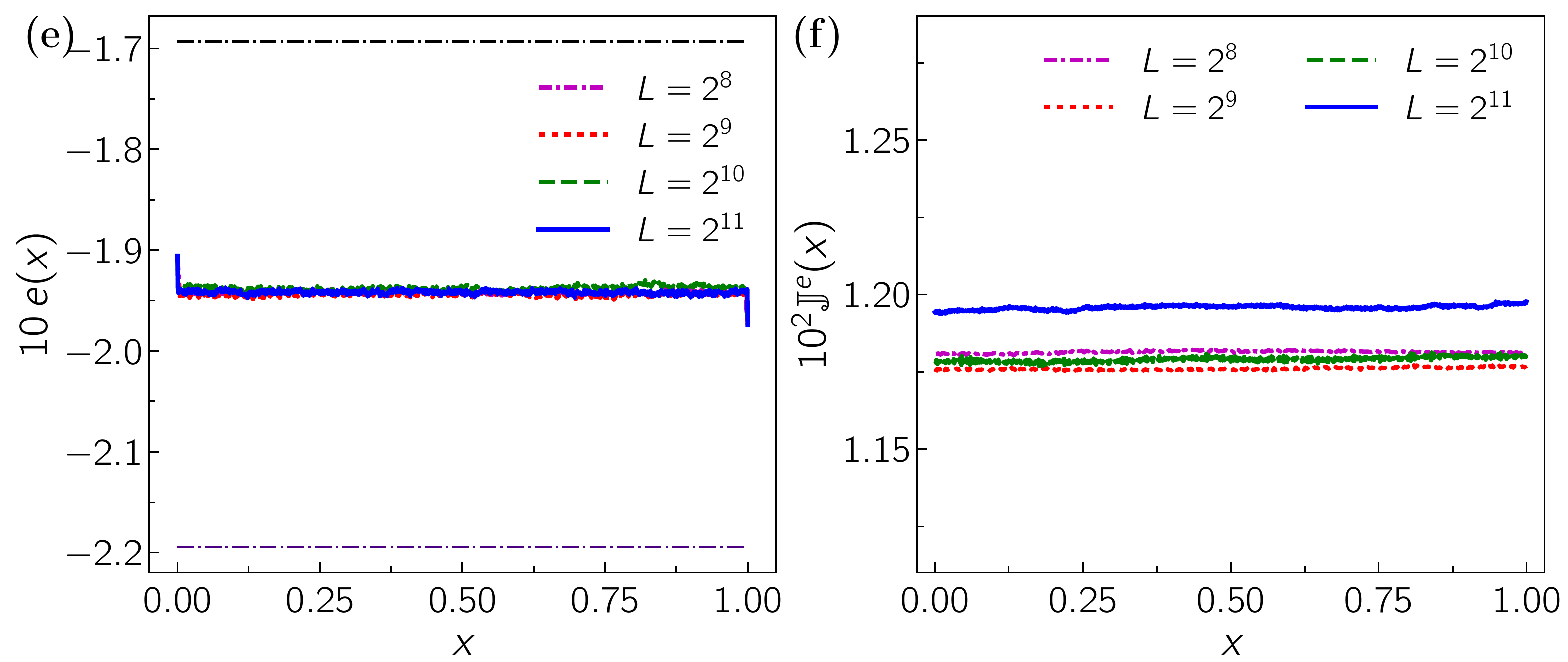}
	\end{subfigure}
	%\end{center}
	\caption{Plots of bond energy profiles in (a), (c) and (e) for the CH I, CH II, and IH models. The parameters for all these cases are provided in Table~\ref{tab:params}. The temperature is $T=0.2$ for the CH I model whereas the temperature is fixed at $T=1$ for CH II and IH models. For CH I and CH II, we observe linear profiles. This indicates diffusive transport for CH I and CH II. On the other hand we observe flat profile for the IH model which suggests ballistic transport. As in \figref{spro-ib}, the horizontal dashed-dotted lines show expected values of bond energy at the boundaries. In other words, the top horizontal line shows the expected bond energy at the left boundary while the bottom line shows the expected bond energy at the right boundary. As before, in order to benchmark the numerical procedure and to ensure we have truly reached NESS in the right column [(b), (d), and (f)], we have shown the local average bond current. The flat profiles indeed confirm that we have reached NESS.
	}
	\label{fig:epro-pure}
\end{figure}

\begin{figure}
	\begin{center}
		\begin{subfigure}{1\linewidth}
			\includegraphics[width=1.0\linewidth]{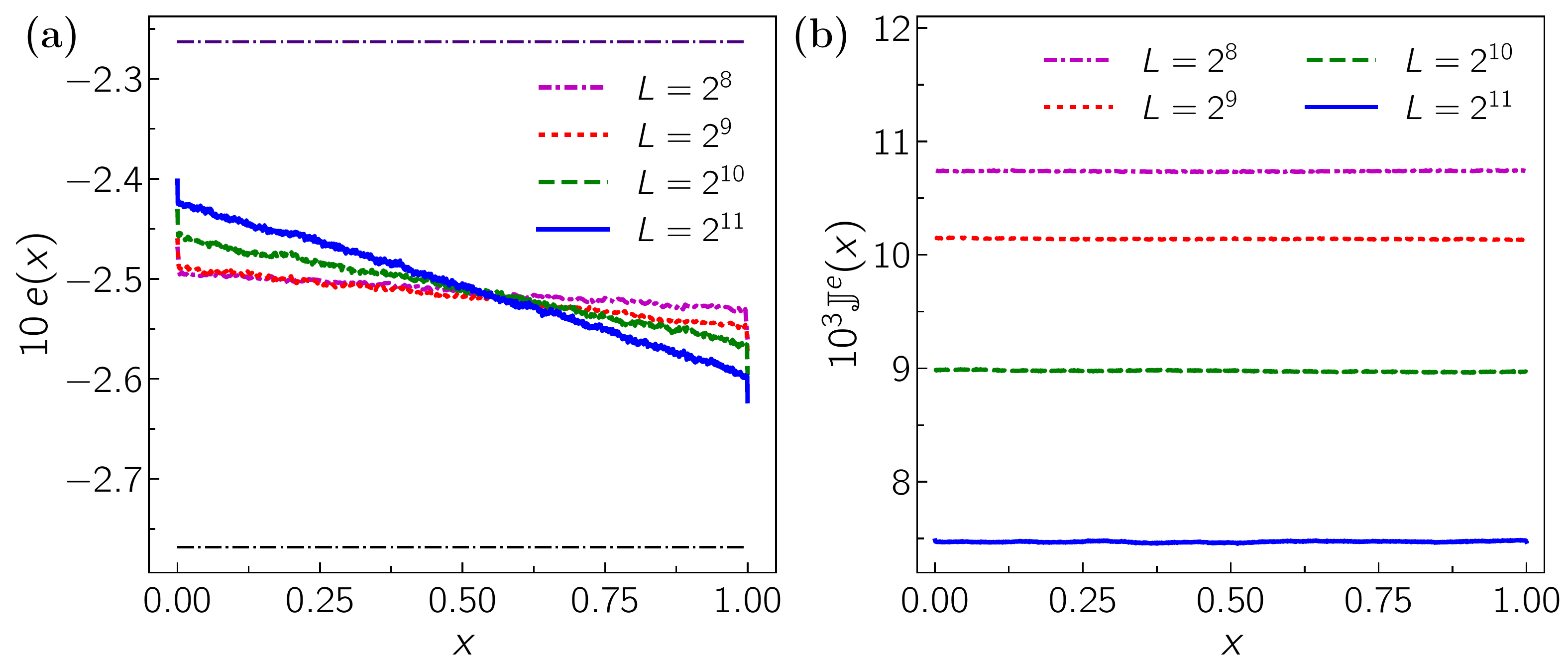}
		\end{subfigure}
		\begin{subfigure}{1\linewidth}
			\includegraphics[width=1.0\linewidth]{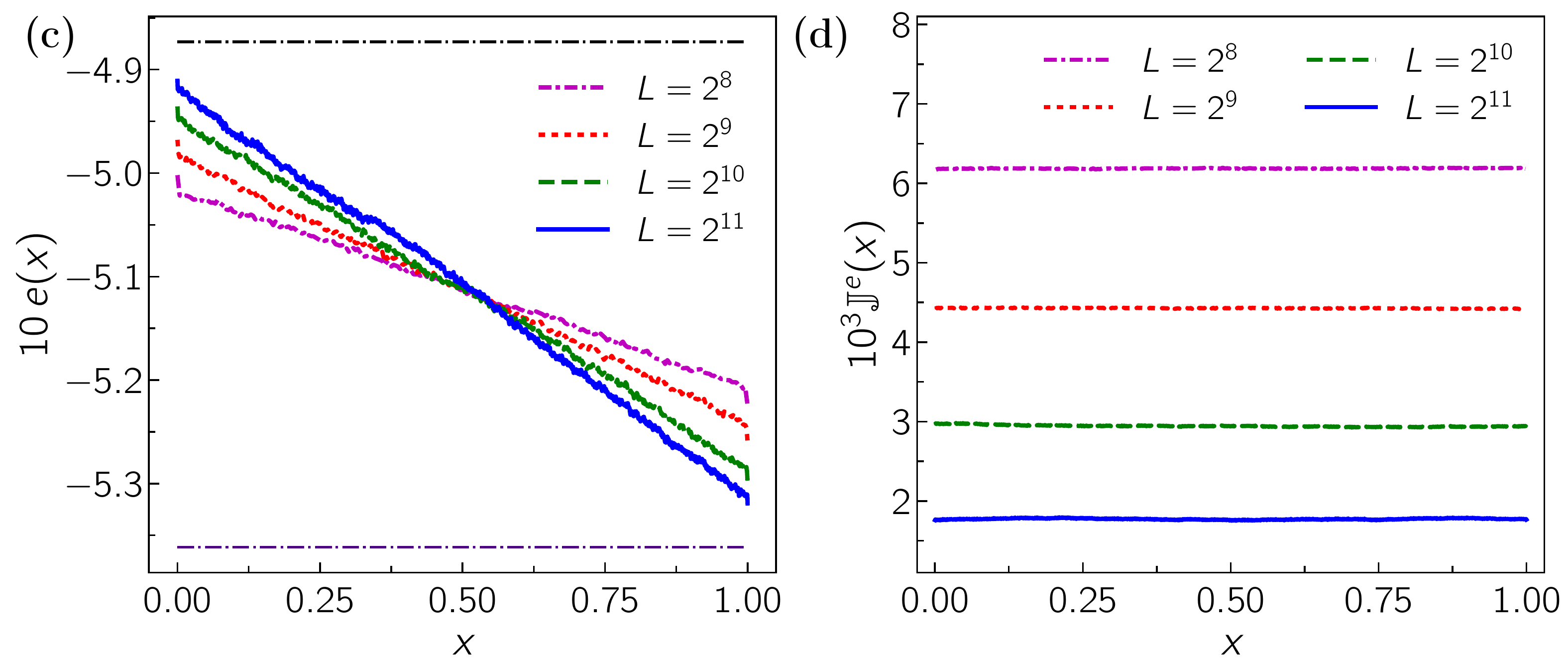}
		\end{subfigure}
		\begin{subfigure}{1\linewidth}
			\includegraphics[width=1.0\linewidth]{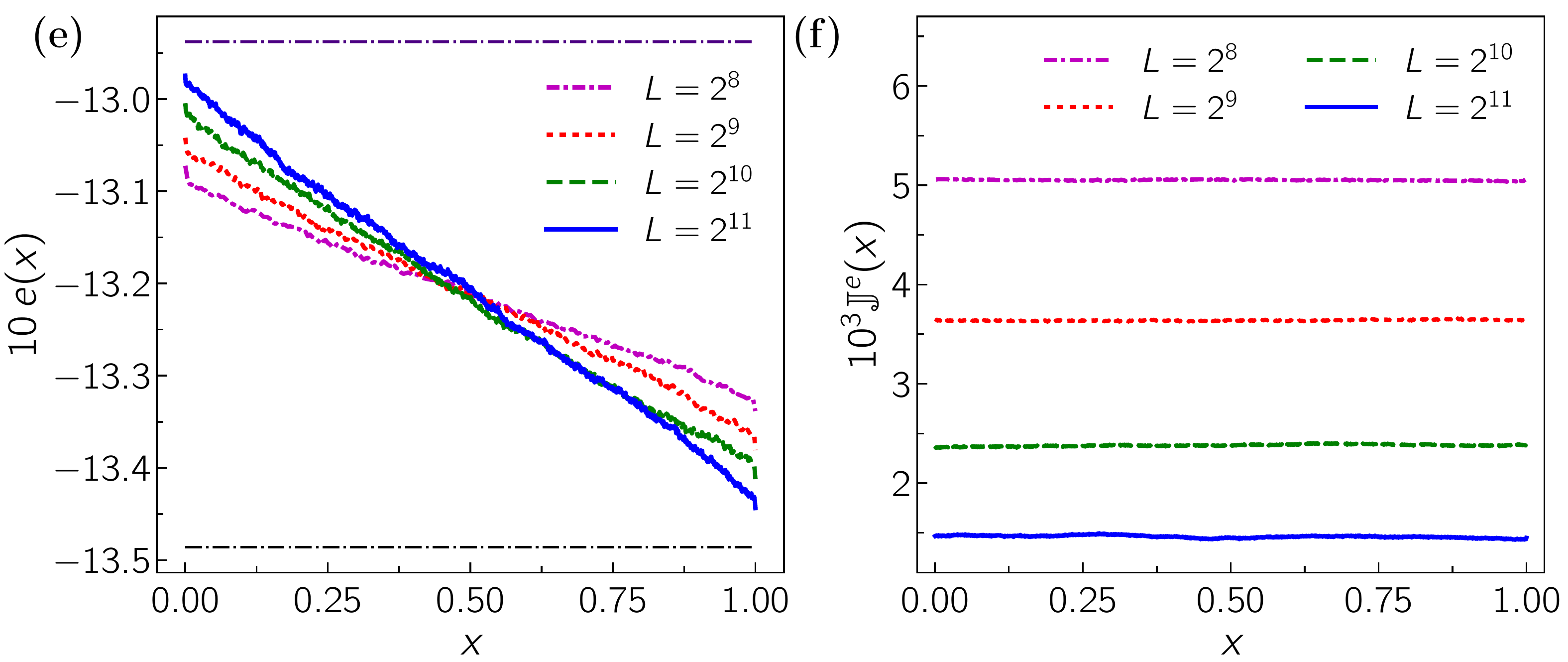}
		\end{subfigure}
		\begin{subfigure}{1\linewidth}
			\includegraphics[width=1.0\linewidth]{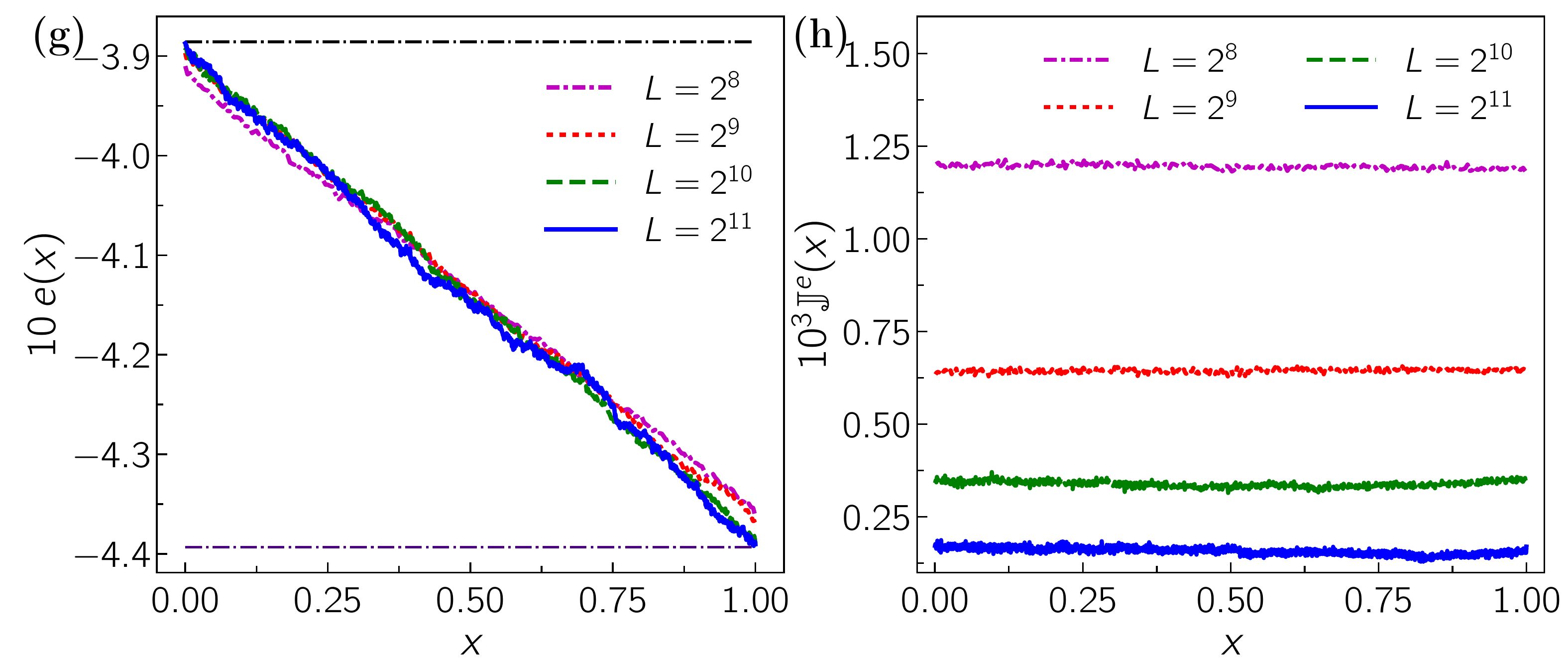}
		\end{subfigure}
	\end{center}
	\caption{Plots of bond energy profiles in (a), (c), (e) and (g) for the integrability-broken cases IH-CH I, II, III, and IV respectively. The parameters for all these cases are provided in Table~\ref{tab:params}. For all these cases, we observe linear profiles. This indicates diffusive transport. As in \figref{epro-pure}, the horizontal dashed-dotted lines show expected values of bond energy at the boundaries. In other words, the top horizontal line shows the expected bond energy at the left boundary while the bottom line shows the expected bond energy at the right boundary. As before, in order to benchmark the numerical procedure and to ensure we have truly reached NESS in the right column [(b), (d), (f) and (h)], we have shown the local average bond current. The flat profiles indeed confirm that we have reached NESS.
	}
	\label{fig:epro-ib}
\end{figure}

\textit{Initial configuration preparation. --} First, we generate a thermal configuration for the bulk spins ($\vec{S}_1 , \vec{S}_2, \ldots, \vec{S}_L$) at temperature $T$ using Monte Carlo (MC) algorithm. In doing so, we consider the entire Hamiltonian in Eq.~\eqref{eq:ham_app}, but update only spins $\vec{S}_1 , \vec{S}_2, \ldots, \vec{S}_L$. Recall that the bulk is defined solely as $\vec{S}_1 , \vec{S}_2, \ldots, \vec{S}_L$. We now discuss the boundary spins ($\vec{S}_0$ and $\vec{S}_{L+1}$). To study the energy transport, we fix the temperature of the left ($\vec{S}_0$) and right ($\vec{S}_{L+1}$) boundary spins as $T_{l} = T + \Delta T $ and $T_{r} = T - \Delta T $ respectively. Keeping $\vec{S}_1$ and $\vec{S}_L$ fixed at the values obtained by Monte Carlo (MC), we separately do an MC simulation to update $\vec{S}_0$ and $\vec{S}_{L+1}$ by considering terms in Hamiltonian in Eq.~\eqref{eq:ham_app} but involving only the bond term $\vec{S}_0 \cdot\vec{S}_1$ and $\vec{S}_{L} \cdot \vec{S}_{L+1}$ respectively. Until now, we have discussed the preparation of an initial state, i.e. a configuration $\vec{S}_0, \vec{S}_1, \ldots, \vec{S}_{L+1}$. The next step is the time evolution. 

%bond terms of the type $\vec{S}_1 \cdot \vec{S}_{2}$ and  $\vec{S}_L \cdot \vec{S}_{L+1}$. The bulk is defined solely as $\vec{S}_1 , \vec{S}_2, \ldots, \vec{S}_L$. 

%and suitable thermal configurations for the boundary spins using Monte Carlo Metropolis algorithm. To study energy transport, we fix the temperature  at  at the left boundary and  at the right boundary with $\Delta T > 0$. We set $T = T_l = T_r$ in the case of spin transport. Moreover, we consider auxiliary fields $\vec{g}_l \text{ and } \vec{g}_r$ at the left and right boundaries (respectively) with 
%\begin{equation}
%	\vec{g}_l = \brac{ 0, 0, g }, \quad \vec{g}_r = \brac{ 0, 0, - g }, \quad g \geqslant 0,
%\end{equation}
%in this case. These fields make the spin at the left (resp. right) boundary point along positive $z$-direction (resp. negative $z$-direction). We use Metropolis Monte Carlo (MC) algorithm to generate the initial condition with these parameters. 

\textit{Time evolution. --} Starting from the initial configuration prepared in the manner discussed above at $t=0$, we evolve \emph{all} spins ($\vec{S}_0, \vec{S}_1, \ldots, \vec{S}_{L+1}$) via deterministic Hamiltonian dynamics using a fourth order adaptive Runge-Kutta method (ARK4) until a time (say denoted by $t_1$) in the simulation becomes just above a chosen threshold time denoted by $t_{up}$. At $t_1$, single MC step is performed at each of the boundary spins ($\vec{S}_0$ and $\vec{S}_{L+1}$ using only the bond terms $\vec{S}_0\cdot \vec{S}_1$ and $\vec{S}_{L}\cdot \vec{S}_{L+1}$ repectively). If the steps of either of the spins are successful, we update the boundary spins accordingly. On the other hand, however, we do not change the bulk spins. With this updated spin configuration (boundary spins) at $t=t_1$, we repeat the deterministic dynamics of the whole chain $\vec{S}_0, \vec{S}_1, \ldots, \vec{S}_{L+1}$, and subsequently do the MC update of the boundary spins. This entire process is repeated for a long time so that the system reaches the NESS. We denote this time as $\mathcal{T}_{ss}$. At this time the bond currents are independent of bond index (see, for example, Fig.~\ref{fig:epro-pure} to be discussed later) which is an indication of successfully reaching NESS.

It is important to note that, Runge-Kutta methods do not preserve conserved quantities. In particular, the lengths of the spins may differ from unity especially if $\mathcal{T}_{ss}$ is too large (for example, say $ > 5 \times 10^5$). Evolving the spins furthermore is expected to incur considerable errors especially in quantities, such as the averaged currents which are generally small ($ \ll 1$) for the chosen parameters of interest. To circumvent this issue, we normalize all the spins at $t=\mathcal{T}_{ss}$. Even though this is an approximation, the system will still be very close to the desired NESS.

Having reached NESS, we still need to repeat the above prescription in order to compute time-averaged currents. 
We start by carrying out the dynamics for a short interval of time (say $\mathcal{T}_{sh}$) just to ensure that we are in the NESS. We then run the simulation for a time $\mathcal{T}_{av}$ after which we record the data. This process is itself performed a number of times (say $4-10$) to get average value of the data.

Until now, for simplicity, we have restricted our discussions to  the case  where only the boundary temperatures are specified. We now briefly outline the extension of the procedure to the case with specified magnetic fields at the two ends. In that case we set (i) $\Delta T =0$ and (ii) consider two  boundary magnetic fields $\vec{g}_{l}$ and $\vec{g}_{r}$ at the boundary spins $\vec{S}_0$ and $\vec{S}_{L+1}$ respectively.
% given by
% \begin{equation}
%     \vec{g}_{0} = \brac{ 0, 0, g }, \quad \vec{g}_{L+1} = \brac{ 0, 0, - g }, \quad g > 0.
% \end{equation}
In order to update the boundary spins $\vec{S}_0$ and $\vec{S}_{L+1}$ using MC, we consider the additional energy contribution of these fields $ -\vec{g}_{l} \cdot \vec{S}_{0} $ and $ -\vec{g}_{r} \cdot \vec{S}_{L+1} $ respectively. This ensures that any desired magnetization is incorporated at the boundaries.

\section{Equilibrium correlations}
Recall that we computed the correlation function $C_{neq}(n)$ defined as
\begin{equation}
    C_{neq}(n) =  \langle \vec{S}_{ \frac{L}{2} } \cdot \vec{S}_{\frac{L}{2} + n} \rangle  - \langle \vec{S}_{\frac{L}{2}} \rangle  \cdot \langle \vec{S}_{\frac{L}{2} + n} \rangle.
    \label{eq:cneq}
\end{equation}
In this section we discuss some analytical forms of equilibrium correlation function and compare our numerics with it. For this purpose it is useful to define the following correlation for the open ($C_{ob}$) and periodic ($C_{pb}$) chains
\begin{eqnarray}
        C_{ob}(n) & = & \langle \vec{S}_{L/2} \cdot \vec{S}_{L/2 + n} \rangle_{eq} 
        \label{eq:cob} \\
        C_{pb}(n) & = &  \frac{1}{L}\sum_{l=1}^{L} \big\langle\vec{S}_{l} \cdot \vec{S}_{l + n} \big\rangle_{eq} ,
        \label{eq:cpb}
\end{eqnarray}
where $\langle \cdot \rangle_{eq}$ is average in equilibrium. In the case of open boundaries, we employ the procedure in Section~\ref{sec:sim-det} with $\Delta T =0$ and $g=0$. This enables us to reach equilibrium and we then average over time to compute $C_{ob}(n)$. On the other hand, we compute $C_{pb}(n)$ by averaging over equilibrium configurations of the periodic chain generated by MC. Next, we discuss some analytical forms for the equilibrium correlation functions in the Ishimori-Haldane model (Section~\ref{ssec:ih-cor}) and the Classical Heisenberg model (Section~\ref{ssec:ch-cor}).

%For the purpose of verification, we compute the correlation in the open chain
%\begin{equation}
%	C_{ob}(n) = \langle \vec{S}_{L/2} \cdot \vec{S}_{L/2 + n} \rangle_{ss} ,
%\end{equation}
%where $\langle \cdot \rangle_{ss}$ denotes average over both ensemble and time in the steady state. In order to compare with the equilibrium correlations, we will set the same temperature throughout the system ($T_l = T = T_r $) and apply zero boundary field. In order to compare, we also compute the correlation $C_{pb}(n) = \langle \sum_{l=1}^{L}\vec{S}_{l} \cdot \vec{S}_{l + n} \rangle_{eq} / L$ using MC in a periodic chain of $L$ spins. The results are shown in \figref{coreq}. The exact results for the CH and IH models are known as we describe below. The simulation results in \figref{coreq} agree well with the analytical values. 

\subsection{Analytical expressions for the two-point correlation function in the the Ishimori-Haldane model}
\label{ssec:ih-cor}

We show here that 
\begin{equation}
    C_{pb}(n) =	\langle \vec{S}_{a} \cdot \vec{S}_{a+n}\rangle_{eq} = \frac{ 1 }{3^ {| n |}}, \quad n \in \mathbb{Z}, 
	\label{eq:ihcor}
\end{equation}
and this holds true in the infinite system-size limit $ L \rightarrow \infty $. Recall that the Hamiltonian for the IH model with $L$ three-component spins is given by
\begin{equation}
	\mathcal{H} = - \sum_{k=1}^{L} \ln \brac{ 1 + \vec{S}_{k} \cdot \vec{S}_{k+1}} \ , 
\end{equation}
where we assume periodic boundary condition $ \vec{S}_{L+1} = \vec{S}_1 $ and set inverse temperature $\beta=1$. To prove \eqref{eq:ihcor}, we write the spins $\vec{S}_{k}$ as 
\begin{equation}
	\begin{aligned}
		S_k^x  & = \sin \theta_k \cos \phi_k, \\
		S_k^y  & = \sin \theta_k \sin \phi_k, \\
		S_k^z  & = \cos \theta_k, \\
	\end{aligned}
\end{equation}
where $\phi_k \in [0, 2 \pi)$ and $\theta_k \in [0, \pi]$. We introduce the notations:
\begin{subequations}
	\begin{align}
		\int \mathcal{D} S    & \equiv \int \cdots \int \prod_{k=1}^{L} \text{d} \vec{S}_{k}  \ , \\	
		\int \text{d} \vec{S}_{k}  & \equiv \int_{0}^{ \pi } \sin{\theta_k} \, \text{d} \theta_k  \int_{0}^{ 2 \pi } \text{d} \phi_k  \ , \\
		p_{kl} & \equiv \vec{S}_k \cdot \vec{S}_{l} .
	\end{align}
\end{subequations}
Thus we have
\begin{equation}
		p_{kl} =  \sin \theta_k  \sin \theta_l  \cos \left( \phi_k - \phi_l \right) + \cos \theta_k  \cos \theta_l . 
\end{equation}
Note that
\begin{subequations}
	\begin{align}
		\int \mathrm{d}\vec{S}_l \ p_{kl} & = 0  \\
		\int \mathrm{d}\vec{S}_l \ p_{kl}^2 & = \frac{16 \pi^2 }{ 3 }  \\
		\int \mathrm{d}\vec{S}_l \ p_{kl}^3 & = 0 \\
		\int \mathrm{d}\vec{S}_l \ p_{kl} \, p_{lm} & = \frac{4 \pi }{ 3 } p_{km}.
	\end{align}
\end{subequations}
Using these relations, we find that the partition function $\mathcal{Z}_{L}$ at $\beta=1$ is given by 
\begin{equation}
    \mathcal{Z}_{L} = \int \mathcal{D} S \prod_{k=1}^{L} \big[1+p_{k(k+1)} \big] = (4 \pi )^{L} + 3 \brac{ \frac{4 \pi}{3} }^L .
    \label{eq:zl-all}
\end{equation}
Note that in Eq.~\eqref{eq:zl-all}, the notation $p_{k(k+1)}$ stands for $\vec{S}_k \cdot \vec{S}_{k+1}$ and Eq.~\eqref{eq:zl-all} holds for any $L$. In addition to this normalization factor $\mathcal{Z}_{L}$ given in Eq.~\eqref{eq:zl-all}, we also need to evaluate the following integral
\begin{equation}
	\mathcal{N}_L (n) = \int \mathcal{D} S \ p_{a(a+n)} \prod_{k=1}^{L} \big[1+p_{k(k+1)} \big] .
\end{equation}
When $n=0$, we have $\mathcal{N}_L(0)=\mathcal{Z}_{L}$. For $|n| \geqslant 1$, the terms which survive on expanding the product are 
\begin{equation}
	\begin{aligned}
		\int \mathcal{D} S \ p_{a(a+n)} \prod_{k=a}^{a+n} p_{k(k+1)} &=  (4 \pi )^L \frac{1 }{3^{|n|}}  \ , \\		
		\int \mathcal{D} S \ p_{a(a+n)} \prod_{k=a+n}^{L+a} p_{k(k+1)} &=  (4 \pi )^L \frac{ 1 }{3^{L-|n|}} . 
	\end{aligned}
\end{equation}
Thus the correlation $\langle \vec{S}_{a} \cdot \vec{S}_{a+n}\rangle_{eq}$ is given by 
\begin{equation}
    \begin{aligned}
        \langle \vec{S}_{a} \cdot \vec{S}_{a+n}\rangle_{eq}  
        & = \frac{\mathcal{N}_L (n) }{ \mathcal{Z}_{L}} \\
        & = \frac{ \frac{1 }{3^{|n|}} + \frac{ 1 }{3^{L-|n|}}  }{  1 + 3 \brac{ \frac{ 1 }{3} }^L }
    \end{aligned}
\end{equation}

Assuming $|n| \ll L$ and taking the limit $L \rightarrow \infty$, we find 
\begin{equation}
	\langle \vec{S}_{a} \cdot \vec{S}_{a+n}\rangle_{eq} =  \frac{\mathcal{N}_{L}(n)}{\mathcal{Z}_{L}} = \frac{1}{3^{|n|}}.
    \label{eq:ih-analytical}
\end{equation}
In \figref{coreq}, we compare the analytical results provided in Eq.~\eqref{eq:ih-analytical} with the direct numerics of the IH model and observe that the numerical results agree well with the analytical results.

\subsection{Exact formulas for equilibrium two-point correlations in the CH model in absence of field}
\label{ssec:ch-cor}
In this subsection, we recall the exact formulas for the equilibrium correlations in the CH model \cite{1967-joyce, 2012-bagchi-mohanty}. Recall that the CH model with $L$ three-component spins is given by 
\begin{equation}
	\mathcal{H} = - \sum_{k=1}^{L} \vec{S}_{k} \cdot \vec{S}_{k+1}
\end{equation}
It turns out that the two-point correlation is 
\begin{equation}
	\langle \vec{S}_{a} \cdot \vec{S}_{a+n}\rangle_{eq} = \left[ \mathcal{L}(1)\right]^n
\end{equation}
where the \emph{Langevin} function $\mathcal{L}(x)$ is given by
\begin{equation}
	\mathcal{L}(x) = \coth(x) - \frac{1}{ x }.
    \label{eq:ch-analytical}
\end{equation}
In \figref{coreq}, we compare the analytical results provided in Eq.~\eqref{eq:ch-analytical} with the direct numerics of the CH model and find that the agreement is excellent.

\begin{table}[htbp!]
    \renewcommand*{\arraystretch}{1.6}
    \bigskip
    \begin{center}
			\begin{tabular}{|c|c|c|c|c|c|c|}
				\hline 
				Model  & Case & \ \ $J$ \ \ & \ \ $\lambda$ \ \ & \ \ $T$ \ \ & \ \ $\Delta T$ \ \ & \ \ $ g $ \ \  \\ \hline
				\multirow{2}{*}{ \ \ CH \ \ } & I & $0$ & $1$ & $0.2$ & $0.025$ & $0.01 $  \\ \cline{2-7}
				& II  & $0$ & $1$ &  $1$  & $0.1$ & $0.1$  \\ \cline{1-7}
				\multirow{1}{*}{IH/ILLL } & & $1$ & $0$ & $1$ & $0.1$ & $0.1$   \\ \cline{1-7}
				\multirow{4}{*}{IH-CH}& I & $1$ & $0.1$ & $1$ & $0.09$ & $0.08$  \\ \cline{2-7}
				& II & $1$ & $0.5$ & $1$ & $0.06$ & $0.05$  \\ \cline{2-7}
				& III & $1$ & $1.5$ & $1$  & $0.04$ & $0.05$  \\ \cline{2-7}
				& IV & $0.2$ & $1$ & $1$ & $0.07$ & $0.08$  \\ 
				\hline
			\end{tabular}
    \end{center}	
    \hspace{0cm}
    \caption{The values of the parameters $T$, $\Delta T$ and $g$ used in DNS. Recall that the auxiliary fields $\vec{g}_l$ and $\vec{g}_r$ are set in terms of the parameter $g$ as $\vec{g}_l = \brac{0,0,g}$ and $\vec{g}_r = \brac{0,0,-g}$ as mentioned in the main text. The values of $\Delta T$ were chosen to keep the energy difference of the left and the right reservoirs approximately at $ 0.05$. The values of $g$ are chosen such that the absolute value of magnetization at the boundaries is close to $ 0.06$. The column ``Case'' labels the parameter set.} 
    \label{tab:params}
\end{table}

\section{Profiles in NESS}
\label{sec_app_profiles}
In this section, we discuss the the spatial profiles of the local magnetization and energy. The spatial profiles can indicate the nature of nonequilibrium transport. For e.g., a linear spatial profile is indicative of conventional diffusive transport and a nonlinear profile is often a fingerprint of anomalous transport. A horizontally flat profile is a sign of ballistic transport behaviour. Below we discuss these spatial profiles. 
%We note that the corresponding current profiles are flat. This confirm that we have indeed reached NESS in our DNS. 
For all cases discussed in this section, the values of parameters, such as the bulk temperature $T$, boundary temperatures ($T \pm \Delta T$), and the strength of boundary fields ($g$), are provided in Table~\ref{tab:params}.

First, we describe the local magnetization $m(x)$ (defined in the main text) which we recall to be 
\begin{equation}
m(x) = \langle S^{z}_{ [xL] } \rangle\, , 
\end{equation}
where $[xL] $ equals the integer closest to $xL$ and recall that $\langle \cdot \rangle$ denote average over time (in NESS) as well as different realizations. In \figref{spro-ib}~(a), (c), (e), and (g), we show the magnetization profiles for the four integrability-broken cases. For all cases, the magnetization profile is nonlinear. This nonlinearity is a hallmark of anomalous (KPZ) transport. In the case of IH-CH III [\figref{spro-ib}~(e)] where $J=1, \lambda=1.5$, we are in the non-perturbative regime and the nonlinear nature is not yet clearly seen. The apparent linear profile for the magnetization profile in this case is potentially a finite-size effect. We expect to obtain a nonlinear profile when the system size $L$ is higher. In order to benchmark the numerical procedure and to ensure we have truly reached NESS in the right column of \figref{spro-ib} [i.e., (b), (d), (f), and (h)], we have shown the local average bond spin current. The flat profiles confirm that we have reached NESS.
%pin current profile is flat throughout the system in all the cases.

We next discuss the spatial energy profiles. The local energy $e_n$ is defined in the main text which we recall to be 
\begin{equation}
e_n  = - J  \ln \! \left( 1 + \vec{S}_n \cdot \vec{S}_{n+1} \right) - \lambda  \vec{S}_n \cdot \vec{S}_{n+1}  
	\label{eq:ham_app_2}
\end{equation}
with $J, \lambda \geqslant 0$ describing the relative strength of the integrable and non-integrable parts. We show the spatial energy profiles in \figref{epro-pure} for the CH and IH models. The energy profile is linear for the CH model at both low and high temperatures [see Figs.~\ref{fig:epro-pure}~(a) and (c)]. 
%We observe convergence of the energy profiles with increase in $L$ in the case of high temperature. For low temperatures, we observe a trend that the energy profiles would converge for higher $L$. 
This linear profiles indicate diffusive behaviour for the CH model. In the IH model, the energy profile is flat for all $L$ [see \figref{epro-pure}~(e)]. This is a signature of ballistic transport. We show the energy profile for the integrability-broken cases in \figref{epro-ib}~(a), (c), (e), and (g). The profiles are linear. However, the profiles are observed to converge with the system size clearly only for the case IH-CH IV [see \figref{epro-ib} (g)]. This hints at diffusive behaviour for the case IH-CH IV. For other cases, we expect that the profiles would converge with system size as we increase system sizes to larger values.  As before, in order to benchmark the numerical procedure and to ensure we have truly reached NESS in the right column of both \figref{epro-pure} and \figref{epro-ib}, we have shown the local average bond energy current. The flat profiles indeed confirm that we have reached NESS.

\bigskip
\bigskip 

\bibliography{sneq-refs.bib}